\documentclass{ieeeaccess}
\usepackage{cite}
\usepackage[colorlinks,urlcolor=blue,linkcolor=blue,citecolor=blue]{hyperref}
\usepackage{amsmath,amssymb,amsfonts}
\usepackage{graphicx}
\usepackage{textcomp}
\usepackage{amsthm}
\usepackage{algorithm}
\usepackage{braket}
\usepackage{algpseudocode}
\usepackage{enumitem}
\usepackage[caption=false,font=footnotesize]{subfig}
\usepackage{array}
\usepackage{tabularx}
\usepackage{color}

\xdef\xfigwd{0pt}

\def\BibTeX{{\rm B\kern-.05em{\sc i\kern-.025em b}\kern-.08em
    T\kern-.1667em\lower.7ex\hbox{E}\kern-.125emX}}
\begin{document}
\history{Date of publication xxxx 00, 0000, date of current version xxxx 00, 0000.}
\doi{10.1109/TQE.2020.DOI}

\title{Treewidth-Aware Gate Cut Selection for Reducing Transpilation Overhead on Superconducting Quantum Devices}
\author{\uppercase{Hana Ebi}\authorrefmark{1},
\uppercase{Shin Nishio\authorrefmark{1,3}, and Takahiko Satoh}.\authorrefmark{2}
}
\address[1]{Graduate School of Science and Technology, Keio University, Yokohama, Kanagawa 223-8522 Japan}
\address[2]{Faculty of Science and Technology, Keio University, Yokohama, Kanagawa 223-8522 Japan
\address[3]{Department of Physics \& Astronomy, University College London, Gower St, London, WC1E 6BT, United Kingdom}
}
\tfootnote{This paper is based on results obtained from a project,
JPNP23003, commissioned by the New Energy and Industrial Technology Development Organization (NEDO).
SN is also supported by the JSPS Overseas Research Fellowship.
TS is also supported by JST Grant Number JPMJPF2221.}

\markboth
{Hana Ebi \headeretal: Treewidth-Aware Gate Cut Selection for
Reducing Transpilation Overhead on
Superconducting Quantum Devices}
{Hana Ebi \headeretal: Treewidth-Aware Gate Cut Selection for
Reducing Transpilation Overhead on
Superconducting Quantum Devices}

\corresp{Corresponding author: Hana Ebi (e-mail: hebi0713@keio.jp).}

\begin{abstract}
On superconducting quantum devices with sparse qubit connectivity, transpilation of long-range two-qubit interactions inserts additional SWAP gates, increasing hardware cost and execution error. Gate cutting via quasi-probability decomposition (QPD) can remove a selected two-qubit gate and thereby reduce routing overhead, but its sampling cost makes cut placement critical. We propose TW2S, a graph-only two-stage gate-cut selection method that operates on the circuit interaction graph without backend-specific transpilation at selection time. Stage 1 analyzes a min-fill elimination trace and scores edges by their contribution to a treewidth upper bound. Stage 2 ranks the resulting candidates by edge betweenness centrality with a degree penalty to identify routing bottlenecks. Across grid, Watts-Strogatz, barbell, and stochastic block model benchmarks transpiled to IBM’s FakeSherbrooke backend, TW2S consistently outperforms random cut selection when the interaction graph contains identifiable sparse cuts. The advantage is governed not by absolute graph density but by moderate community structure and accessible inter-community edges. We further derive a mean-squared-error breakeven condition showing that, under a shared total shot budget, QPD is beneficial only when the ECR reduction is large enough and the signal strength is sufficient. Under an expanded per-subcircuit budget the signal-strength requirement is substantially relaxed. In noisy simulations of the J1-J2 transverse-field Ising model, TW2S achieves $\Delta$ECR = 47 for n = 8, compared with approximately 9 for random selection, and yields lower estimation error than the uncut baseline in the tested strong-signal regime, with larger gains at increased shot budgets. These results position graph-structural cut selection as a practical compiler-side tool for turning circuit cutting into a targeted routing-reduction strategy.
\end{abstract}

\begin{keywords}
Quantum circuit cutting, Quasi-probability decomposition, Routing overhead reduction, Superconducting quantum devices, Treewidth heuristic
\end{keywords}

\titlepgskip=-15pt

\maketitle

\section{Introduction}
\label{sec:introduction}

\PARstart{S}{uperconducting} quantum processors are among the leading platforms
for near-term quantum computing, but their sparse hardware
connectivity makes the execution of the non-local two-qubit
interactions expensive~\cite{supreconducting}.
When a quantum circuit is mapped to a nearest-neighbor device, such as a heavy-hex, the
transpiler must insert additional SWAP gates to bring distant
qubits together.
Because implementing each SWAP requires multiple native two-qubit gates,
routing overhead can substantially increase the two-qubit gate count, often becoming the dominant source of
execution error on the current superconducting
hardware~\cite{quantumcircuit_optimization,wille2019mapping,nishio2020extracting}.
Reducing this routing-induced two-qubit overhead is therefore a
central problem in quantum compilation.

An alternative strategy is circuit cutting.
By replacing a selected two-qubit gate with a quasi-probability
decomposition~(QPD), one can split the circuit into smaller
subcircuits that may be routed more
independently~\cite{peng2020,mitarai2021}.
Recent gate-cutting-based virtual-gate demonstrations on
superconducting devices suggest that this idea can suppress errors
by removing costly long-range
interactions~\cite{mitarai2021,Yamamoto2023VTQG}.
The difficulty, however, is that cutting also introduces sampling
overhead.
A useful cut must therefore do more than simply partition the
circuit. It must remove an interaction whose absence substantially
reduces post-transpilation two-qubit cost.

\begin{figure}[htbp]
\centering
\includegraphics[width=\linewidth]{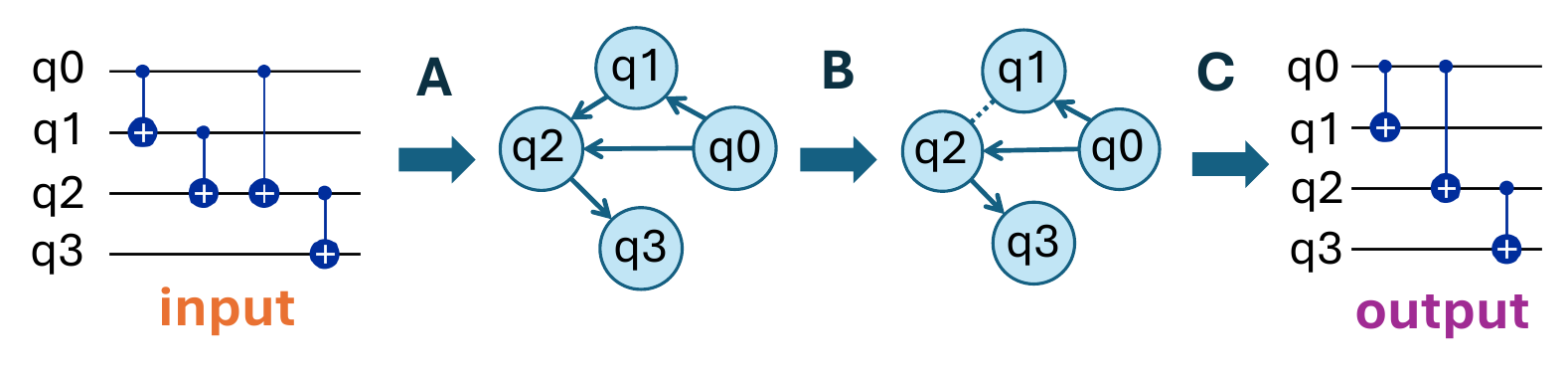}
\caption{Overview of the proposed TW2S cut selection pipeline.
(A) A quantum circuit is represented as an interaction graph, where nodes correspond to qubits and edges correspond to two-qubit gates.
(B) TW2S selects the cut position in two stages. Stage 1 uses a min-fill treewidth ordering to identify candidate edges that lie on sparse inter-community boundaries, and Stage 2 ranks these candidates by edge betweenness centrality with a degree penalty to select the final cut.
(C) The selected edge is cut via quasi-probability decomposition (QPD), splitting the original circuit into smaller sub-circuits that can be executed independently on a noisy device and recombined to estimate the target observable.}
\label{fig:overview}
\end{figure}

This observation motivates a transpiler-side cut-selection problem:
identifying which single two-qubit gate to cut so as to reduce
routing overhead most effectively.
In this work, we formulate this problem in graph-theoretic terms.
We represent the circuit as an interaction graph and identify
candidate cut edges from structural bottlenecks in that graph.
Our method, TW2S, proceeds in two stages.
Stage~1 analyzes a min-fill elimination trace and scores edges by
how strongly they contribute to a treewidth upper bound, thereby
shortlisting structurally costly interactions.
Stage~2 ranks these candidates using edge betweenness centrality
with a degree penalty, favoring edges that act as genuine routing
bottlenecks rather than hub-adjacent local connections.
Because both stages operate only on graph-structural quantities,
TW2S avoids backend-specific transpilation calls at selection
time.
Figure~\ref{fig:overview} provides an overview of the proposed
framework.

The contributions of this paper are threefold.
First, we propose a graph-only two-stage selector for single-gate
QPD cutting that targets routing overhead reduction rather than
qubit-count reduction.
Second, we evaluate TW2S on grid, Watts--Strogatz, barbell, and stochastic block model graph families.
These evaluations show that TW2S outperforms random cut selection when the interaction graph contains identifiable sparse cuts.
Moreover, we characterize the regime in which TW2S provides the strongest benefit in terms of moderate community structure.
Third, using a mean-squared-error breakeven analysis and noisy
simulations of the \mbox{$J_1$--$J_2$} transverse-field Ising model on
\textsc{FakeSherbrooke}, we
clarify when two-qubit gate reduction translates into improved estimation
accuracy and when QPD fails because of mid-circuit measurement
overhead or weak signal strength.
Taken together, these results position gate cutting not only as a
tool for decomposing large circuits, but also as a practical
method for reducing routing overhead on sparse superconducting
hardware when the circuit structure and signal regime are
favorable.

\section{Background}
\label{sec:background}
\subsection{Quantum Circuit Transpilation and Qubit Routing}
\label{subsec:transpilation}

A quantum circuit written at the abstract level does not impose constraints on the connectivity between qubits.
On a physical superconducting device, however,
two-qubit gates can be applied only between qubits that are
directly connected in the device's \emph{coupling graph}.
For instance, IBM's heavy-hex lattice provides a sparse
coupling graph with maximum degree three~\cite{Chamberland2020heavyhex}.

When a logical circuit contains a two-qubit gate on a non-adjacent qubit pair, a typical compiler inserts additional SWAP gates to bring the two qubits into proximity. Each SWAP gate decomposes into three native two-qubit gates~\cite{NielsenChuang2010} (e.g., the echoed cross-resonance (ECR) gate on IBM Quantum processors), so even a single SWAP insertion substantially increases the two-qubit gate count.
This process of mapping logical qubits to physical qubits and inserting SWAPs to satisfy connectivity constraints is collectively called \emph{transpilation}~\cite{quantumcircuit_optimization}.

The qubit routing problem (deciding how to map and move logical
qubits throughout execution) is NP-hard in
general~\cite{Siraichi2018QubitA, Cowtan2019routing}, so practical
compilers rely on heuristic methods.
A range of heuristic routers have been proposed, including
$A^*$-based search~\cite{zulehner2018efficient} and
architecture-aware passes such as
t$|$ket$\rangle$~\cite{Cowtan2019routing}.
In this work all transpilations are performed with the
widely-adopted SABRE algorithm~\cite{10.1145/3297858.3304023} at
Qiskit~\cite{Javadi2024qiskit} optimization level~1. SABRE
greedily inserts SWAPs to reduce the distance between operands of
pending two-qubit gates.

Because two-qubit gate errors dominate the overall
execution error on current superconducting
devices~\cite{supreconducting, Krantz2019guide, Tannu2019}, the post-transpilation count
of native two-qubit gates (specifically ECR gates on the
backend used in this study) serves as a
hardware-relevant proxy for execution cost.
Reducing this count is the primary objective of the
gate-cut selection method proposed in
Section~\ref{sec:method}.

\subsection{Gate Cutting via Quasi-Probability Decomposition}
\label{subsec:qpd}

Circuit cutting~\cite{peng2020,mitarai2021,mitarai2021overhead} is a family of
techniques that decomposes a quantum circuit into smaller
subcircuits whose measurement results are recombined
through classical post-processing.
In this work we employ \emph{gate cutting} exclusively,
in which a selected two-qubit gate is replaced by
local operations on independent subcircuits.

In gate cutting based on quasi-probability decomposition
(QPD)~\cite{mitarai2021}, a single CX (or CZ) gate is
decomposed into $6$ subcircuit pairs.
Each pair is assigned a real coefficient
$c_k\in\{+\tfrac{1}{2},-\tfrac{1}{2}\}$, and the
expectation value of an observable $O$ for the original
circuit is reconstructed as
\begin{equation}
  \label{eq:qpd-reconstruction}
  \langle O \rangle
  = \sum_{k=1}^{6} c_k\, \langle O \rangle_k,
\end{equation}
where $\langle O \rangle_k$ is the expectation value obtained
from the $k$-th subcircuit pair.
Because the coefficients include negative values, the
reconstruction is not a convex combination.
This introduces a sampling overhead: for a single CX/CZ
cut, the overhead factor is
$\gamma = \sum_k |c_k| = 3$, so the variance of the
reconstructed estimator increases by a factor of
$\gamma^2 = 9$ relative to direct execution with the same
number of shots per subcircuit.
For $K$ simultaneous cuts the overhead grows as
$\gamma^{2K}$~\cite{mitarai2021overhead}, making it essential to keep $K$ small.

In the present study we fix $K=1$ and exploit gate cutting
not to partition a circuit across multiple devices, but to
remove a single routing-costly two-qubit gate from the
circuit prior to transpilation.
If the removed gate corresponds to a long-range interaction
that would otherwise require multiple SWAP insertions, the
resulting subcircuits can be transpiled with significantly
fewer native two-qubit gates.

\subsection{Treewidth and Routing Overhead}
\label{subsec:treewidth}

Given a quantum circuit, we define its \emph{interaction graph}
$I=(V_I,E_I)$, where $V_I$ is the set of logical qubits and
an edge $\{u,v\}\in E_I$ exists whenever a two-qubit gate acts
on the pair $(u,v)$.
Similarly, the target device is described by its
\emph{coupling graph} $C=(V_C,E_C)$.

A standard measure of tree-likeness is the treewidth $\mathrm{tw}(G)$.
A tree has treewidth~$1$, and graphs with more complex connectivity
have larger treewidth~\cite{ROBERTSON198449,ROBERTSON1986309,BODLAENDER1998}.
Intuitively, a small treewidth means the graph can be embedded into
a sparse device with little routing overhead.
This motivates selective gate cutting to reshape the interaction
graph by removing structurally critical edges, as described in
Sec.~\ref{sec:method}.

Computing treewidth exactly is NP-hard in
general \cite{doi:10.1137/0608024}, but useful upper bounds can
be obtained efficiently via elimination heuristics.
Given an ordering $\pi=(v_1,\dots,v_n)$ of the vertices, we
define the associated elimination process as follows. At step
$i$, the vertex $v_i$ is eliminated after adding fill edges so
that all of its remaining neighbors form a clique. The
elimination width $w(\pi)$ is the maximum number of remaining
neighbors of $v_i$ at the time of its elimination. 

It is well known that $\mathrm{tw}(G)\le w(\pi)$ for any
ordering~$\pi$ \cite{Rose1976AlgorithmicAO,BODLAENDER1998}.
In this work, we use the \emph{min-fill}
heuristic \cite{BODLAENDER2010259}, which greedily selects the
vertex whose removal introduces the fewest fill edges, to
obtain an elimination ordering and the associated upper
bound.

Importantly, the elimination process provides more than a
coarse upper bound on treewidth. The bag sizes and the number
of fill edges introduced during elimination also reveal where
the interaction graph locally departs from a tree-like
structure. Such locations can be interpreted as structural
bottlenecks that are likely to induce routing overhead after
transpilation. 
\section{Breakeven Analysis of QPD Gate Cutting}
\label{sec:breakeven_analysis}
In this section, we examine \emph{when} gate
cutting can be expected to improve estimation accuracy at all.
A theoretical breakeven condition under an idealized
depolarizing-bias model can be derived (full derivation in
Appendix~\ref{app:theoretical_breakeven}). Here we summarize
its essentials and then identify the practical conditions
under which this idealized criterion breaks down on realistic
noisy hardware (Sec.~\ref{sec:failure_modes}).
The resulting conditions guide both the design of TW2S
(Sec.~\ref{sec:method}) and the interpretation of the noisy
simulations in Sec.~\ref{sec:noisy_sim}.
\paragraph{Idealized breakeven model}
Throughout this section we adopt an idealized model with
(i) a single QPD cut ($K_\mathrm{cut} = 1$);
(ii) a depolarizing-like bias model in which the expectation-value
bias depends only on the post-transpilation ECR count $N$ via
$\mathrm{bias}(N) \approx
|\langle H\rangle_\mathrm{ideal}|(1 - e^{-pN})$;
(iii) noiseless mid-circuit operations introduced by QPD;
(iv) a fixed observable variance $\sigma_H^2$ unaffected by cutting.
The QPD sampling overhead is $\gamma^2 = 9$ per cut
(Sec.~\ref{subsec:qpd}). We denote the post-cut ECR-count
reduction by $\Delta N$, the same quantity referred to as
$\Delta\mathrm{ECR}$ in the empirical sections
(Sec.~\ref{sec:algo_verification} onwards).
 
Modelling MSE as the sum of squared bias and variance and
comparing $\mathrm{MSE}_{\mathrm{QPD}}$ against
$\mathrm{MSE}_{\mathrm{base}}$ yields, for the shared-budget
($\times 1$) strategy, the breakeven shot count
\begin{equation}
  M^*(\Delta N)
  = \frac{(\gamma^2-1)\,\sigma_H^2}
         {|\langle H\rangle_{\mathrm{ideal}}|^2
          \!\left[\left(1-e^{-pN}\right)^2
                 -\left(1-e^{-p(N-\Delta N)}\right)^2\right]},
  \label{eq:Mstar}
\end{equation}
above which QPD with equal total budget wins.
For the per-subcircuit $\times 1.5$ ($\times 9$ total) strategy,
the variance penalty is exactly cancelled and gate cutting is
advantageous whenever $\Delta N > 0$ in this idealized model.
The three key levers in Eq.~\eqref{eq:Mstar} are:
\begin{enumerate}
  \item \textbf{$\Delta N$}: the primary output of
        TW2S. Larger reduction drives $M^*$ down.
  \item \textbf{$|\langle H\rangle_{\mathrm{ideal}}|$}: lowers $M^*$
        quadratically. As it approaches zero, $M^* \to \infty$ and
        QPD cannot win at any finite shot count.
  \item \textbf{$pN$}: most favourable near unity (circuits deep
        enough to accumulate significant but not saturated noise).
\end{enumerate}
The detailed derivation, the $\times 9$ analysis, and a
visualisation of $M^*(\Delta N)$ for representative parameters
are deferred to Appendix~\ref{app:theoretical_breakeven}.

\subsection{Failure-Mode Analysis}
\label{sec:failure_modes}
 
Neither the $\times 9$ guarantee nor the $\times 1$ breakeven
condition holds universally on realistic hardware.
To characterize the practical deviations, we conduct a systematic
noisy-simulation experiment on the one-dimensional transverse-field
Ising model (1D TFIM) benchmark circuit,
sweeping over circuit size, Trotter depth, shot budget, and two
shot-allocation strategies ($\times 1$ shared and $\times 9$
per-subcircuit).
The full configuration grid and circuit parameters are given in
Appendix~\ref{app:failure_mode_params}.

\begin{figure*}[t]
  \centering
  \includegraphics[width=\linewidth]{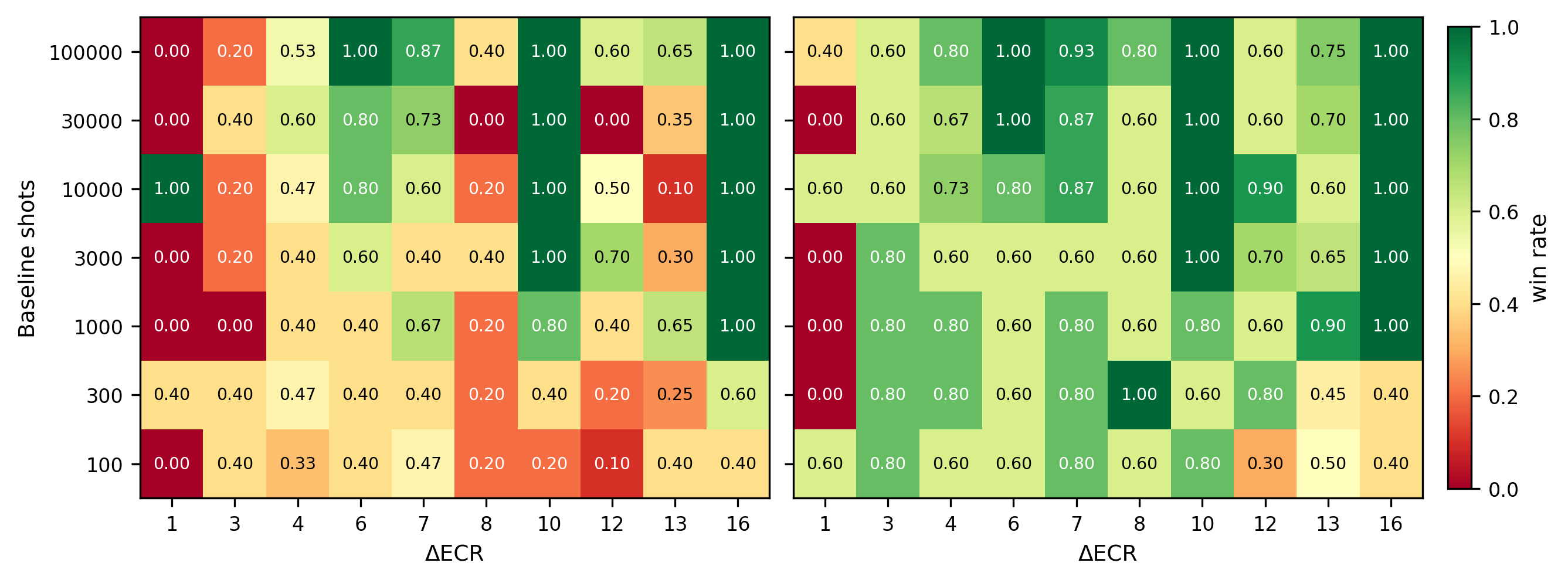}
  \caption{Empirical QPD win rate in the
    $(\Delta\mathrm{ECR},\, M)$ plane,
    restricted to strong-signal configurations
    ($|\langle H\rangle_{\mathrm{ideal}}| > 0.5$).
    Left: shared ($\times 1$) strategy.
    Right: per-subcircuit $\times 1.5$ ($\times 9$ total) strategy.
    Each cell aggregates over all configurations sharing the same
    $\Delta\mathrm{ECR}$, with $R = 5$ repetitions per configuration.
    Green = QPD wins; red = baseline wins.}
  \label{fig:heatmap_strategy}
\end{figure*}

\begin{figure*}[t]
  \centering
  \includegraphics[width=0.9\linewidth]{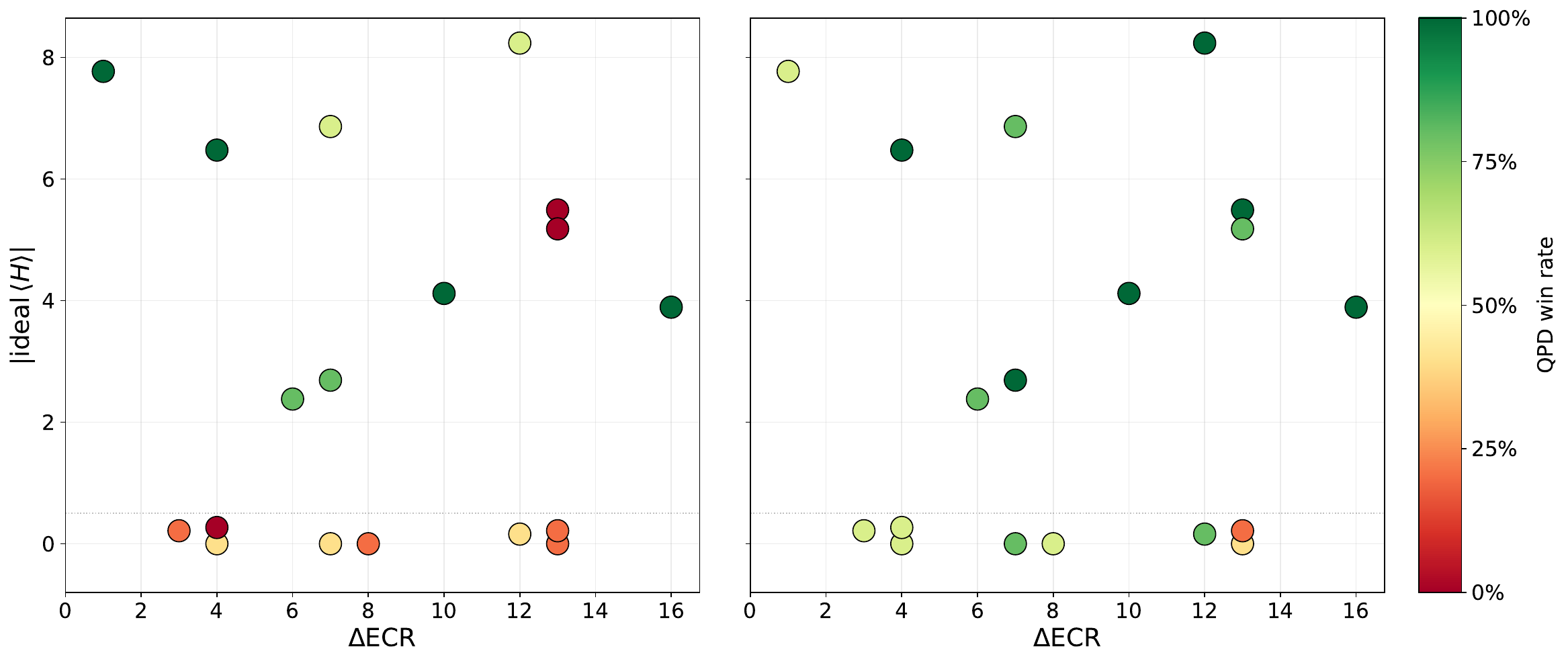}
  \caption{QPD win rate at $M = 10{,}000$ shots in the
    $(\Delta\mathrm{ECR},\;|\langle H\rangle_{\mathrm{ideal}}|)$
    plane for shared ($\times 1$, left) and per-subcircuit
    $\times 1.5$ ($\times 9$, right) strategies.
    Color encodes win rate (green = QPD wins, red = baseline wins).
    Odd-step configurations cluster near
    $|\langle H\rangle_{\mathrm{ideal}}| \approx 0$ (bottom),
    even-step configurations at
    $|\langle H\rangle_{\mathrm{ideal}}| \gtrsim 2.4$ (top).}
  \label{fig:scatter_shots10k}
\end{figure*}

Two primary failure mechanisms emerge from this sweep.
Figure~\ref{fig:heatmap_strategy} shows the empirical win rate of
QPD in the $(\Delta\mathrm{ECR},\, M)$ plane for both strategies,
restricted to configurations with
$|\langle H\rangle_{\mathrm{ideal}}| > 0.5$ (even Trotter steps)
to isolate the effect of ECR reduction and shot count on
Failure Mode~1.
Figure~\ref{fig:scatter_shots10k} shows all 18 configurations in
the $(\Delta\mathrm{ECR},\;|\langle H\rangle_{\mathrm{ideal}}|)$
plane at $M = 10{,}000$ shots, colored by win rate, exposing
the role of signal strength in Failure Mode~2.
 
\subsubsection{Failure Mode 1: Mid-circuit Measurement Errors}
\label{sec:midcircuit_error}
 
\begin{figure*}[t]
  \centering
  \includegraphics[width=\linewidth]{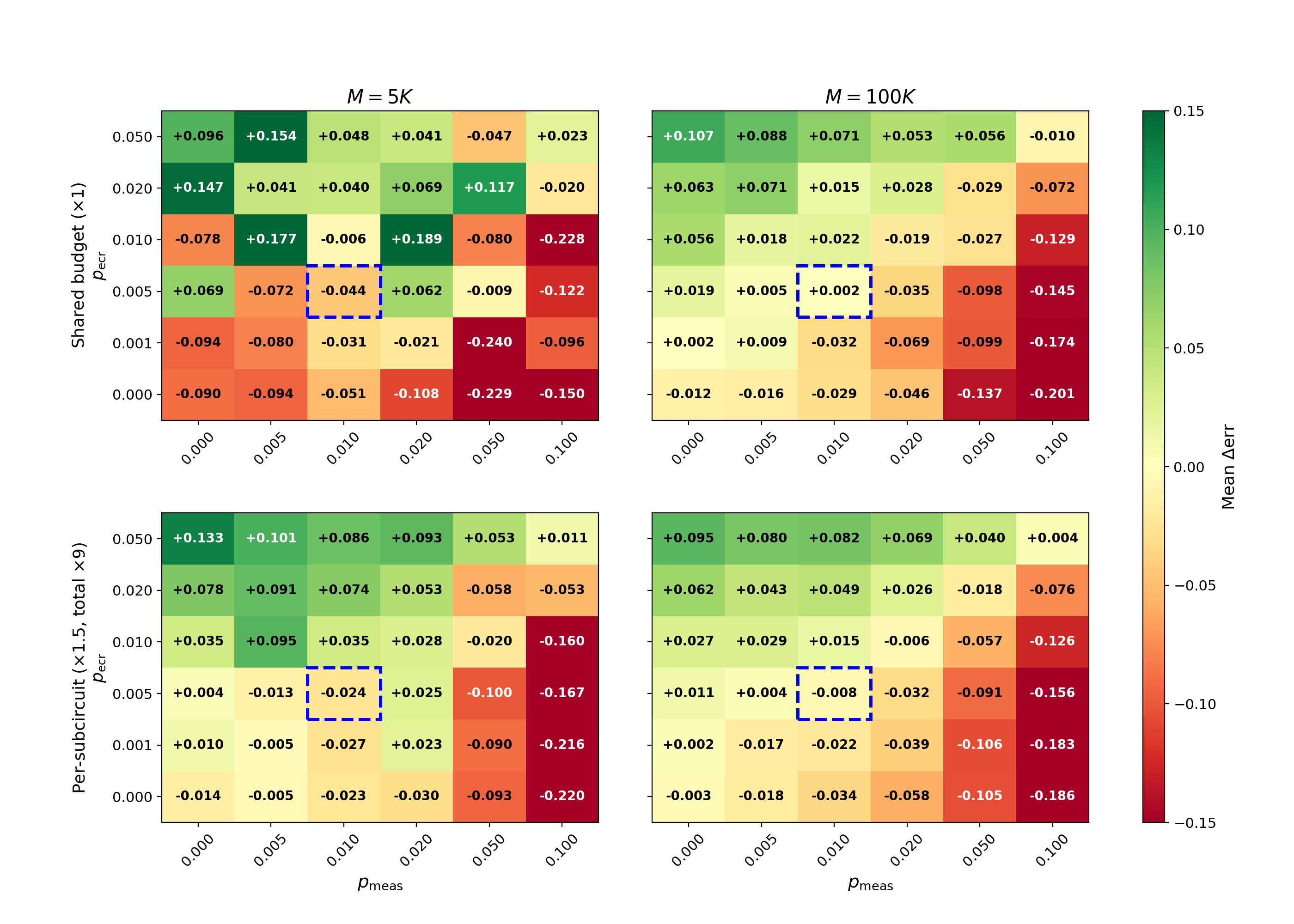}
  \caption{Mean $\Delta\mathrm{err}$ as a function of
    $p_{\mathrm{ecr}}$ (rows) and $p_{\mathrm{meas}}$ (columns),
    for $M \in \{5\mathrm{K},\,100\mathrm{K}\}$ and two strategies
    (shared budget, top; per-subcircuit~$\times 1.5$, bottom).
    Blue dashed box: IBM FakeSherbrooke operating point.}
  \label{fig:pecr_pmeas_heatmap}
\end{figure*}

Equation~\eqref{eq:Mstar} treats QPD mid-circuit operations as
noiseless, but in FakeSherbrooke the mid-circuit measurement error
rate $p_{\mathrm{meas}} \approx 0.01$ exceeds the ECR error rate
$p_{\mathrm{ecr}} \approx 0.005$.
Each QPD cut substitutes one ECR error for a mid-circuit measurement
error. The net bias change per cut is
\begin{equation}
  \Delta\mathrm{bias}_{\mathrm{per\text{-}cut}}
  \approx
  |\langle H\rangle_{\mathrm{ideal}}|
  \!\left(\underbrace{e^{-p}}_{\text{ECR removed}}
        - \underbrace{e^{-p_{\mathrm{meas}}}}_{\text{mid-circuit added}}
  \right),
  \label{eq:bias_exchange}
\end{equation}
which is negligible or negative when $p_{\mathrm{meas}} \ge p$.
A controlled sweep over $(p_{\mathrm{ecr}},\, p_{\mathrm{meas}})$
confirms this crossover near $p_{\mathrm{meas}} \approx p_{\mathrm{ecr}}$
(Figure~\ref{fig:pecr_pmeas_heatmap}).
 
Crucially, this is a \emph{per-cut} overhead that is amortised over
$\Delta\mathrm{ECR}$: empirically, at $\Delta\mathrm{ECR} = 5$ the
per-$\times 1.5$ strong-signal win rate reaches $60\%$ at the IBM
operating point, and the larger $\Delta\mathrm{ECR} = 47$ achieved
by TW2S for $n = 8$ in Section~\ref{sec:noisy_sim} comfortably
exceeds this threshold, which is why TW2S wins robustly there
despite the same $p_{\mathrm{meas}}$.
 
\subsubsection{Failure Mode 2: Weak Signal Strength}
\label{sec:signal_strength}
 
The dominant failure mechanism in the shared-budget strategy is a
small ideal expectation value $|\langle H\rangle_{\mathrm{ideal}}|$.
In our 1D TFIM benchmark, odd Trotter steps ($T \in \{1, 3\}$)
empirically yield small or vanishing ideal expectation values:
$T = 1$ produces $|\langle H\rangle_{\mathrm{ideal}}| \approx 0$
across all $n$, and $T = 3$ produces
$|\langle H\rangle_{\mathrm{ideal}}| \le 0.27$,
while even steps ($T \in \{2, 4\}$) yield
$|\langle H\rangle_{\mathrm{ideal}}| \ge 2.38$
(at least $3.89$ for $n \ge 6$).
This pattern arises from a near-cancellation of the per-site
energy contributions at odd $T$ under the fixed-angle
Trotterisation, and serves here as a natural way to vary
$|\langle H\rangle_{\mathrm{ideal}}|$ across configurations.
 
Figure~\ref{fig:scatter_shots10k} makes this separation visible at
$M = 10{,}000$ shots. Odd-step configurations cluster along the
bottom of each panel ($|\langle H\rangle_{\mathrm{ideal}}| \approx 0$)
and are colored predominantly red under the shared strategy,
while even-step configurations occupy the upper region and are
mostly green.
Quantitatively, odd-step configurations achieve a win rate of only
$25\%$ (shared) and $58\%$ (per-$\times 1.5$) at $M = 10{,}000$
shots, whereas even-step configurations achieve $68\%$ and $90\%$,
respectively.
Crucially, in the shared strategy the odd-step win rate is
substantially lower than the even-step rate, and this separation is
\emph{independent of $\Delta\mathrm{ECR}$}: odd-step configurations
span $\Delta\mathrm{ECR} \in \{3, 4, 7, 8, 12, 13\}$, yet all show
near-zero win rates regardless of how many ECR gates are saved by
cutting.
Under per-$\times 1.5$, by contrast, the cancellation of the
$\gamma^2$ variance penalty rescues many odd-step configurations
(overall odd-step win rate $\approx 58\%$), confirming that the
weak-signal failure is mediated by the variance overhead rather
than by an intrinsic absence of bias reduction.
This behavior can be understood as follows. When
$|\langle H\rangle_{\mathrm{ideal}}| \approx 0$, the absolute bias
reduction obtained by cutting two-qubit gates tends to be limited,
whereas QPD still incurs the $\gamma^2 = 9$ variance overhead in the
shared strategy.
Equation~\eqref{eq:Mstar} captures this exactly: as
$|\langle H\rangle_{\mathrm{ideal}}| \to 0$, $M^* \to \infty$ and
no finite shared-budget shot count makes QPD competitive.
This implies that the applicability of QPD cutting under a shared
budget depends not only on the circuit structure but also on the
physics of the problem being simulated.

\subsection{Conditions for QPD Advantage}
\label{sec:qpd_conditions}
 
Combining both failure modes, QPD reliably outperforms the baseline
when the following conditions hold simultaneously:
\begin{enumerate}
  \item $\Delta\mathrm{ECR}$ is large enough to amortise the per-cut
        mid-circuit measurement overhead
        (empirically, $\Delta\mathrm{ECR} \gtrsim 5$ already
        delivers a substantial win-rate jump at the IBM operating
        point, with TW2S routinely achieving much larger reductions).
  \item $|\langle H\rangle_{\mathrm{ideal}}|$ is sufficiently large
        ($\gtrsim 2$ in our experiments). This condition is
        strictly required only under the shared ($\times 1$)
        strategy and is largely relaxed under per-$\times 1.5$,
        where the variance penalty is cancelled.
\end{enumerate}
Among these, condition~(1) is the only lever directly controllable
by a cut-selection method: the larger the $\Delta\mathrm{ECR}$
achieved at selection time, the more easily condition~(1) is
satisfied and the lower the shot budget required to cross the
breakeven threshold $M^*(\Delta N)$ of Eq.~\eqref{eq:Mstar}.
TW2S is designed to maximize $\Delta\mathrm{ECR}$, the primary
lever in condition~(1), and the noisy simulations of
Section~\ref{sec:noisy_sim} confirm that this translates into
robust error reduction when conditions~(1) and~(2) are both met.
\section{TW2S: A Two-Stage Algorithm for Gate Cut Selection}
\label{sec:method}

\subsection{Problem Setting}
\label{subsec:problem-setting}
 
In this work, we determine which gate to cut in order to minimize routing overhead when exactly one two-qubit gate is removed via QPD
($K=1$).
The input is a quantum circuit $QC$. The output is a single gate $g\in G_{\mathrm{2Q}}(QC)$ selected as the cut target,
where $G_{\mathrm{2Q}}(QC)$ denotes the set of all two-qubit
gates in $QC$.
The selection is based solely on structural properties of
the interaction graph derived from $QC$, without invoking
any backend-specific transpilation at selection time.
The goal is to identify the gate whose removal is expected
to reduce routing overhead most effectively.
 
\paragraph{Interaction graph.}
We extract the two-qubit interaction structure from the input
circuit $QC$ as a preprocessing step.
For each pair of logical qubits $\{u,v\}$ on which at least
one CX or CZ gate acts, we define the \emph{interaction
weight} $w(u,v)$ as the number of such gates.
The interaction graph is $I=(V_I,E_I)$ with
$V_I=\{0,1,\ldots,n{-}1\}$ and
\begin{equation}
  E_I=\bigl\{\{u,v\}\;\big|\; w(u,v)>0\bigr\}.
\end{equation}
We treat $I$ as undirected because the orientation of a
two-qubit gate can generally be absorbed by local
single-qubit transformations, whose cost is negligible compared
with the cost of additional two-qubit gates in our setting.
When the same pair $\{u,v\}$ appears in multiple gate
instances at different positions in the circuit, we also record
the list of their occurrence indices so that a concrete gate
instance can be selected in a later stage.
Because $I$ is constructed from the circuit alone, all
subsequent operations of TW2S are graph-only and do not require
any backend-specific transpilation.

\paragraph{Two-stage selection.}
Our algorithm selects a cut gate in two sequential stages,
each exploiting a different structural property of the
interaction graph $I$.
 
\begin{description}
  \item[Stage~1 (treewidth-guided shortlisting).]
    We run the min-fill elimination heuristic on $I$ and
    score every edge by its contribution to treewidth
    increase (Section~\ref{subsec:bottleneck-ranking}).
    The top-$K$ edges by this score form a shortlist
    $E_{\mathrm{cand}}$.
  \item[Stage~2 (structural tiebreaker).]
    Within $E_{\mathrm{cand}}$, we rank edges by a
    complementary structural score that combines
    edge betweenness centrality and a node-degree penalty
    (Section~\ref{subsec:structural-tiebreaker}).
    The edge with the highest Stage-2 score is selected as
    the cut target.
\end{description}
 
By separating coarse structural filtering (Stage~1) from
fine-grained selection (Stage~2), the algorithm avoids
exhaustive transpilation-based evaluation while remaining
sensitive to both treewidth structure and routing cost.
Algorithm~\ref{alg:two-stage} summarizes the complete
procedure. Its computational cost is analyzed in
Appendix~\ref{app:complexity}.
 
\subsection{Stage 1: Candidate Shortlisting via Min-Fill Trace}
\label{subsec:bottleneck-ranking}
 
As discussed in Section~\ref{subsec:treewidth}, edges whose
removal lowers a treewidth upper bound of $I$ are promising
cut candidates.
To identify such edges efficiently, we run the min-fill
elimination heuristic on $I$ and analyze the resulting trace.
 
\subsubsection{Elimination trace}
The min-fill heuristic constructs an elimination ordering
$\pi$ by greedily removing the vertex that introduces the
fewest fill edges. Before removing each vertex $v$, we record
the \emph{bag} $B(v)=\{v\}\cup N(v)$, where $N(v)$ is the
current neighborhood of $v$. We also record the corresponding
\emph{fill edges}
$F(v)=\bigl\{\{x,y\}\subseteq N(v)
\mid \{x,y\}\notin E(I)\bigr\}$,
which are the edges required to make $N(v)$ a clique.
The maximum bag size over all steps yields a treewidth
upper bound:
$\mathrm{tw}_{\mathrm{ub}}=\max_v(|B(v)|-1)$.
Steps with large $|B(v)|$ and $|F(v)|$ are those that
raise this bound, and the edges incident to such steps are
the structural bottlenecks we wish to target.
 
\begin{figure}[t]
    \centering
    \includegraphics[width=0.9\linewidth]{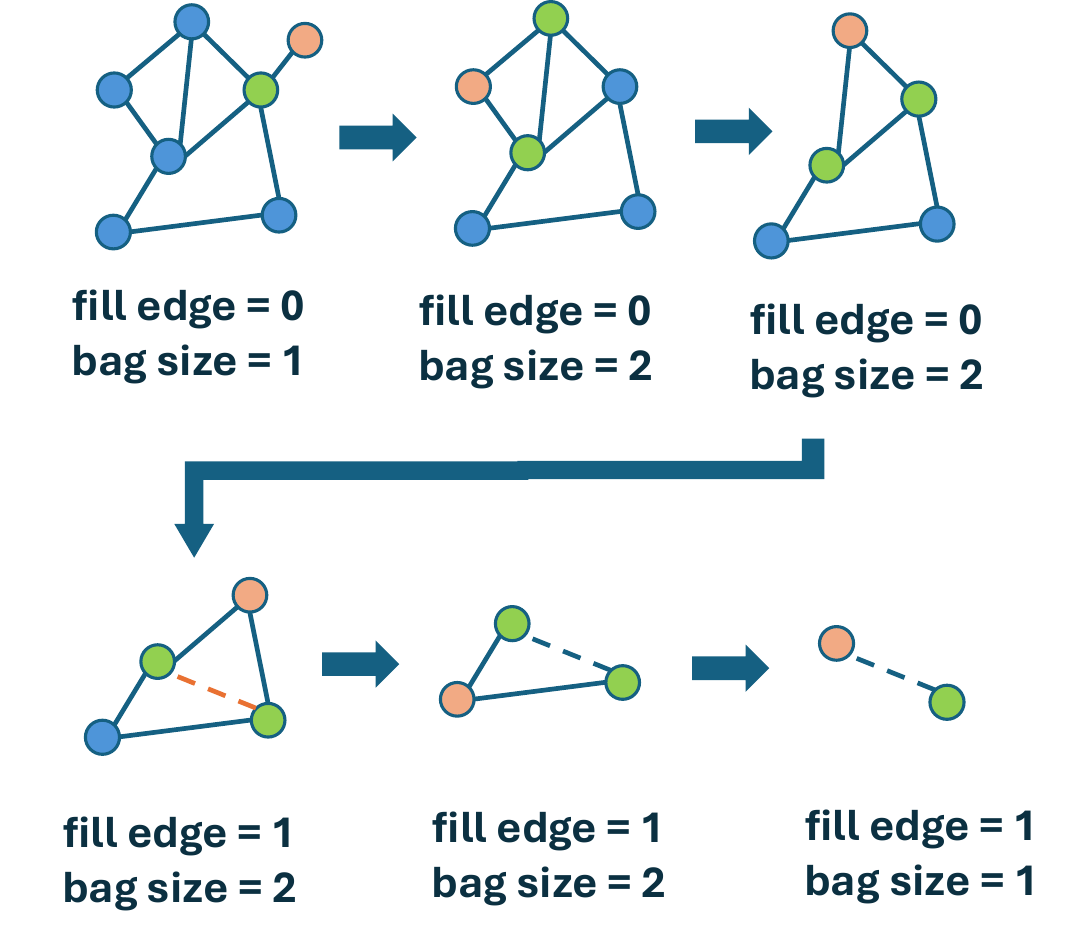}
    \caption{Example of a min-fill elimination trace on an interaction graph.
    At each step, the vertex with the fewest required fill edges, shown as red vertices, is eliminated,
    and missing edges are added among its current neighbors, shown as green vertices, to form a clique.
    The figure illustrates how bags $B(v)=\{v\}\cup N(v)$ and fill edges
    $F(v)$ arise during elimination, which are later used to score candidate
    cut edges in Stage~1.}
    \label{fig:minfill-elimination}
\end{figure}

Figure~\ref{fig:minfill-elimination} illustrates an example of
a min-fill elimination process. At each elimination step, the
selected vertex is removed after adding the minimum number of
fill edges needed to make its current neighborhood a clique.
This trace provides both the bag structure and the fill-edge
information used in Stage~1 scoring.
 
\subsubsection{Edge scoring}
We assign a score to each original interaction edge
$e=\{a,b\}\in E_I$ by attributing responsibility for fill-edge
creation.
When eliminating vertex $v$ produces a fill edge
$\{x,y\}\in F(v)$, the original edges $\{v,x\}$ and
$\{v,y\}$ are regarded as responsible.
Their scores are incremented proportionally to a
step-importance weight
\begin{equation}
  g(v)=\alpha\,(|B(v)|-1)+\beta\,|F(v)|,
\end{equation}
where $\alpha,\beta\ge 0$ are hyperparameters. Throughout this work we set $\alpha=\beta=1$ for simplicity,
giving equal weight to bag size and fill count.
The final score also incorporates the interaction weight
$w(e)$ to prioritize edges that are both structurally important
and frequent in the circuit:
\begin{multline}
  \mathrm{score}_1(e)
  = w(e)\sum_{\text{elim.\ steps }v}
    \Biggl[
      g(v) \\
  \times \sum_{\{x,y\}\in F(v)}
      \mathbf{1}\!\Bigl(
        e\in\bigl\{\{v,x\},\{v,y\}\bigr\}
      \Bigr)
    \Biggr].
\end{multline}
Edges are sorted by $\mathrm{score}_1(e)$ in descending order,
and the top $K$ edges that have at least one gate instance
in the circuit form the shortlist
$E_{\mathrm{cand}}=\{e_1,\ldots,e_K\}$.
 
\subsection{Stage 2: Structural Tiebreaker}
\label{subsec:structural-tiebreaker}
 
Stage~1 narrows the search to $K$ structurally promising
edges, but edges with similar treewidth scores may differ
substantially in their routing cost impact.
Rather than resolving this tie by expensive transpilation-based
ECR evaluation for each candidate, we rank them by a lightweight
score combining two graph-structural features.

The first feature is the \emph{edge betweenness centrality} (BC)
of $e=\{u,v\}$ in $I$,
\begin{equation}
  \mathrm{BC}(e)
  = \sum_{s\neq t\in V_I}
    \frac{\sigma(s,t\mid e)}{\sigma(s,t)},
\end{equation}
where $\sigma(s,t)$ is the total number of shortest paths
from $s$ to $t$ in $I$, and $\sigma(s,t\mid e)$ is the
number of those paths passing through~$e$, normalized to
$[0,1]$ by dividing by the maximum possible value.
Edges with high $\mathrm{BC}$ act as bridges or bottlenecks
in the interaction topology: cutting such an edge partitions
the circuit into subcircuits with fewer long-range
interactions, which tends to reduce routing overhead.
However, hub-adjacent edges can accumulate high $\mathrm{BC}$
without being genuine routing bottlenecks, since the hub
remains densely connected after the cut.
We therefore introduce a \emph{node-degree penalty} (DP)
\begin{equation}
  \mathrm{DP}(e)
  = \frac{\deg_I(u)+\deg_I(v)}{2\,d_{\max}},
\end{equation}
where $d_{\max}=\max_{q\in V_I}\deg_I(q)$, which
down-weights edges whose endpoints participate in many
other interactions.
The Stage-2 score combines these two features as
\begin{equation}
  \mathrm{score}_2(e)
  = \alpha_2\,\mathrm{BC}(e)
    - \beta_2\,\mathrm{DP}(e),
\end{equation}
where $\alpha_2,\beta_2\ge 0$ are hyperparameters
(set to $\alpha_2=1.0$, $\beta_2=0.3$ throughout this work).
Among the $K$ shortlisted candidates, the edge
$e^*=\arg\max_{e\in E_{\mathrm{cand}}}\mathrm{score}_2(e)$
is selected as the cut target.
Because $\mathrm{score}_2$ requires only standard graph
computations, Stage~2 adds negligible overhead compared to
transpilation-based evaluation.

When $e^*$ corresponds to multiple gate instances in $QC$,
we select the first occurrence (smallest circuit-layer index),
as earlier gates are more likely to lie on the critical path
and have a larger effect on routing cost.
The output gate index can be passed directly to any
gate-cutting back-end.
The entire procedure operates on graph-structural quantities
alone, requiring no transpilation calls at selection time.
Its computational cost is summarized in
Appendix~\ref{app:complexity}.
 
\begin{algorithm}[t]
\caption{Two-Stage Cut Selection}
\label{alg:two-stage}
\begin{algorithmic}[1]
\Require circuit $QC$, shortlist size $K$ (default $K=3$),
         hyperparameters $\alpha,\beta,\alpha_2,\beta_2$
\Ensure  gate index $i^*$
\State Compute interaction weights $w(e)$ and gate-index lists for all $e\in E_I$
\State Run min-fill trace on $I$; compute $\mathrm{score}_1(e)$ for all $e\in E_I$
\State $E_{\mathrm{cand}}\leftarrow$ top-$K$ edges in $E_I$ by $\mathrm{score}_1$ with at least one gate instance
\State Compute $\mathrm{BC}(e)$ (normalized) for all $e\in E_{\mathrm{cand}}$ on $I$
\State $e^*\leftarrow\arg\max_{e\in E_{\mathrm{cand}}}
         \bigl(\alpha_2\,\mathrm{BC}(e)-\beta_2\,\mathrm{DP}(e)\bigr)$
\State $i^*\leftarrow$ index of the first gate instance on $e^*$ in $QC$
\State \Return $i^*$
\end{algorithmic}
\end{algorithm}
\section{TW2S Performance Evaluation}
\label{sec:algo_verification}

We evaluate TW2S across four structurally diverse graph families
(Grid, Watts--Strogatz, Barbell, and stochastic block model) to
establish where it outperforms random cut selection and to identify
the structural conditions under which the advantage emerges.
Table~\ref{tab:perf_summary} summarises the per-family results.

\subsection{Experimental Setup}
\label{ssec:ecr_setup}
 
\paragraph{Graph families and circuit construction.}
We evaluate four families:
 
\begin{description}
\item[Grid ($n_\mathrm{rows} \times n_\mathrm{cols}$).]
      Rectangular grids with $n_\mathrm{rows} \in \{3,4,5,6\}$ and
      $n_\mathrm{cols} \in \{3,4,5,6\}$, giving 16 instances with
      $n \in [9, 36]$ qubits.
 
  \item[Watts--Strogatz (WS).]
        Graphs with $n = 20$ nodes, rewiring probability $p = 0.1$,
        and degree $k \in \{2, 4, 6\}$, with 20 random seeds each
        (60 total instances).
        WS graphs model small-world topologies with locally clustered
        structure.
 
  \item[Barbell.]
        Two complete graphs of size $k \in \{3,4,5,6,8,10\}$
        connected by a bridge of length $m \in \{0,1,2,3\}$
        (24 instances total).
 
  \item[Stochastic Block Model (SBM).]
        Two-community graphs with $n = 16$ nodes ($n_\mathrm{per} = 8$,
        $m = 2$), intra-community probability $p_\mathrm{in} = 0.5$,
        and mixing ratio $\mu = p_\mathrm{out}/p_\mathrm{in} \in
        \{0.02, 0.05, 0.10, 0.15, 0.20, 0.30, 0.40\}$, with 20 seeds
        per $\mu$ (140 instances total).
\end{description}
 
For each graph $G$, the quantum circuit is constructed by mapping
each edge $(u,v) \in E$ to a CNOT gate on qubits $u$ and $v$.
This construction ensures that the circuit interaction structure
is identical to the problem graph, allowing clean evaluation of
graph-structural effects on cut performance.
These benchmarks isolate graph-structural effects and are not
intended to model realistic temporal gate multiplicities or
algorithmic circuit ordering, which are addressed separately by
the $J_1$--$J_2$ TFIM evaluation in
Sec.~\ref{sec:noisy_sim}.
 
\paragraph{Backend and transpilation.}
All circuits are transpiled onto \texttt{FakeSherbrooke}, a 127-qubit
IBM Eagle processor with heavy-hex connectivity, using Qiskit's
\texttt{transpile} function at optimization level~1.
To reduce the variance from SABRE's stochastic layout, each circuit
is transpiled with 3 independent random seeds and we report the mean
ECR count.
 
\paragraph{Compared strategies.}
For each circuit instance we compare two strategies:
 
\begin{enumerate}
  \item \textbf{TW2S (proposed).}  The two-stage algorithm of
        Sec.~\ref{sec:method}: Stage~1 selects up to $K = 3$
        candidate edges via min-fill treewidth ordering. Stage~2
        picks the final cut by betweenness centrality with
        degree penalty.
  \item \textbf{Random cut.}  A gate instance is chosen uniformly
        at random from all two-qubit gates, repeated over 5 trials
        and averaged.
\end{enumerate}
 
The \emph{TW2S advantage} is defined as
\begin{equation}
  \Delta_{\mathrm{adv}}
    \;=\; \Delta\mathrm{ECR}_{\mathrm{TW2S}}
    \;-\; \overline{\Delta\mathrm{ECR}_{\mathrm{rand}}},
  \label{eq:adv}
\end{equation}
where $\Delta\mathrm{ECR} = \mathrm{ECR}_{\mathrm{uncut}} -
\mathrm{ECR}_{\mathrm{cut}}$ (positive = fewer gates after cutting).
Statistical significance is assessed by a one-sample $t$-test of
$\Delta_{\mathrm{adv}}$ against zero.
The sample unit for the $t$-test is one graph instance:
$\mathrm{ECR}_{\mathrm{uncut}}$ and $\mathrm{ECR}_{\mathrm{cut}}$
are each averaged over 3 SABRE transpilation seeds
(seeds 42, 123, 7), and $\overline{\Delta\mathrm{ECR}_{\mathrm{rand}}}$
is averaged over 5 independent random-cut trials,
yielding a single $\Delta_{\mathrm{adv}}$ value per instance
before the test.

\subsection{Results}
\label{ssec:ecr_results}
 
\begin{table}[t]
  \centering
  \caption{TW2S advantage ($\Delta_{\mathrm{adv}}$, ECR gates)
           versus random cut selection across graph families.
           Win rate: fraction of instances where TW2S outperforms
           random. $p$-value: one-sample $t$-test of
           $\Delta_{\mathrm{adv}}$ against zero.
           Full per-condition results in Appendix~\ref{app:ecr_per_graph}.}
  \label{tab:perf_summary}
  \begin{tabularx}{\columnwidth}{l >{\raggedright\arraybackslash}X r r l}
    \hline
    Family & Condition & Adv. & Win rate & $p$ \\
    \hline
    Grid    & $3{\times}3$--$6{\times}6$   & $+2.2$  & 69\%  & $0.295$  \\
    WS      & $k=2$, $p=0.1$, $n=20$       & $+6.9$  & 50\%  & $0.029$  \\
    WS      & $k=4$, $p=0.1$, $n=20$       & $+6.4$  & 85\%  & $<0.001$ \\
    WS      & $k=6$, $p=0.1$, $n=20$       & $+1.7$  & 65\%  & $0.382$  \\
    Barbell & $k=3$ (small clique)          & $+1.2$  & 100\% & $0.037$  \\
    Barbell & $k=4$                         & $+1.8$  & 100\% & $0.017$  \\
    Barbell & $k=5$                         & $-3.3$  &  0\%  & $0.011$  \\
    Barbell & $k=8$ (large clique)          & $-10.2$ & 25\%  & $0.120$  \\
    Barbell & $k=10$                        & $+0.4$  & 75\%  & $0.858$  \\
    SBM     & $\mu=0.10$                    & $+4.5$  & 85\%  & $0.003$  \\
    SBM     & $\mu=0.15$                    & $+4.8$  & 80\%  & $<0.001$ \\
    SBM     & $\mu=0.20$                    & $+5.1$  & 85\%  & $<0.001$ \\
    SBM     & $\mu=0.40$                    & $+1.5$  & 50\%  & $0.430$  \\
    \hline
    Overall & all families                  & $+3.3$  & 70\%  & $<0.001$ \\
    \hline
  \end{tabularx}
\end{table}
 
Table~\ref{tab:perf_summary} summarises TW2S performance across
the four families. Full per-family analyses are deferred to
Appendix~\ref{app:ecr_per_graph}.
In brief: on \textbf{Grid} graphs TW2S achieves a positive mean
advantage ($+2.2$ ECR) but the result is not statistically
significant across the full $3{\times}3$--$6{\times}6$ range
($p = 0.295$). On \textbf{Watts--Strogatz} graphs the advantage
is strongest at $k=4$ (85\% win rate, mean $+6.4$ ECR,
$p < 0.001$) and weakens at higher degree. On
\textbf{Barbell} graphs TW2S reliably identifies the bridge
edge for small cliques ($k \in \{3,4\}$, 100\% win rate) but
yields negative advantage for large cliques ($k \geq 5$), where
the bridge is less dominant relative to intra-clique routing.
The remainder of this section focuses on SBM, which serves as the
controlled testbed for the structural conditions analysis in
Sec.~\ref{sec:structural_analysis}.
 
\paragraph{SBM graphs: mixing ratio governs performance.}
The TW2S advantage on SBM graphs depends strongly on the mixing ratio
$\mu = p_\mathrm{out}/p_\mathrm{in}$.
Significant advantage ($p < 0.05$) is observed for $\mu \in [0.07, 0.29]$
across the $p_\mathrm{in} \times p_\mathrm{out}$ sweep,
peaking at $\mu = 0.15$--$0.20$ (mean $+4.8$--$+5.1$ ECR,
win rate 80--85\%, $p < 0.001$).
At $\mu = 0.40$, where community boundaries become diffuse, the
advantage shrinks to $+1.5$ ECR and is no longer statistically
significant ($p = 0.43$).
The structural mechanism underlying this $\mu$-dependence is the
subject of Sec.~\ref{sec:structural_analysis}.
\section{Structural analysis}
\label{sec:structural_analysis}

\subsection{Structural Conditions for TW2S Effectiveness}
\label{ssec:tw2s_conditions}

The per-family results above suggest a unifying pattern: TW2S
performs strongly when the interaction graph possesses an
identifiable \emph{sparse cut} (an edge or small edge set whose
removal maximally decouples the circuit into subgraphs that the
transpiler can route independently).
Using the SBM family as a controlled testbed, we characterize
the structural conditions under which TW2S is effective.

\begin{figure}[h]
  \centering
  \includegraphics[width=1\linewidth]{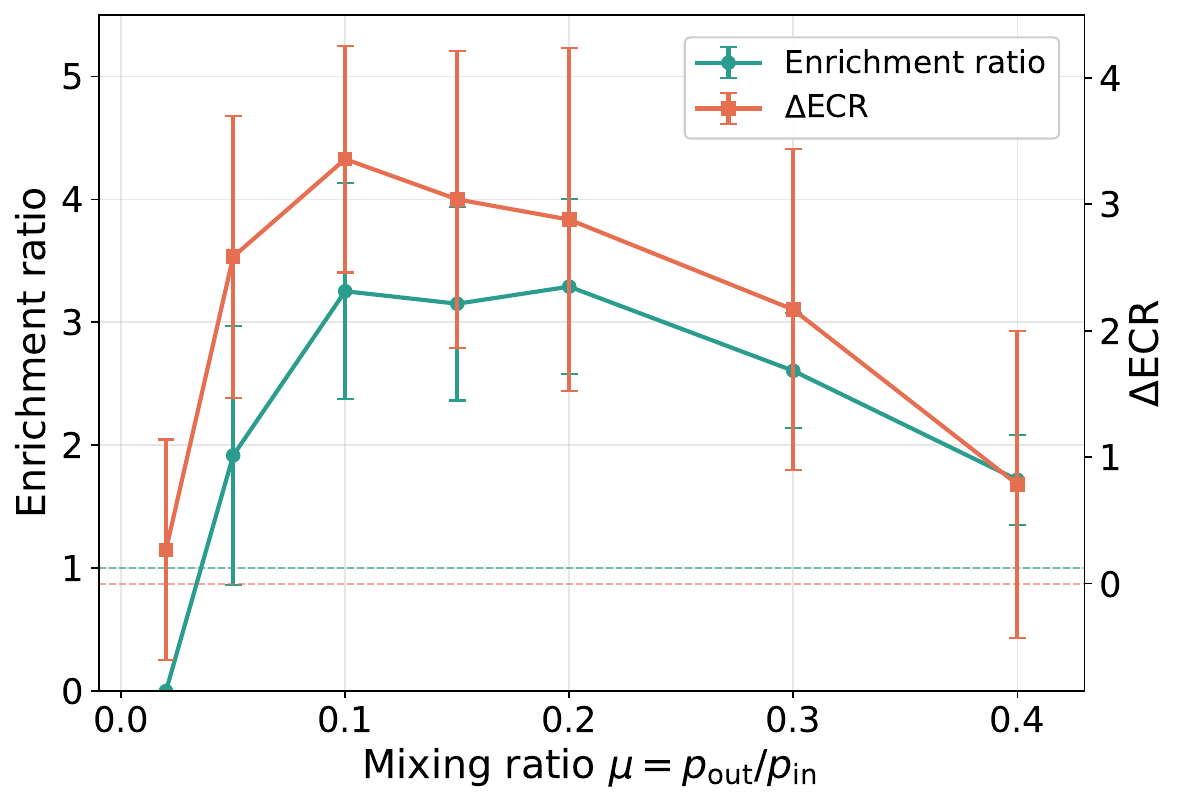}
  \caption{Mixing ratio $\mu$ versus enrichment ratio (left axis,
           teal) and Stage-1 (TW-1) advantage (right axis, orange)
           for SBM graphs ($n=16$, $m=2$, 20 seeds per $\mu$).
           The enrichment ratio peaks near $\mu = 0.10$--$0.15$, coinciding
           with the highest Stage-1 advantage.
           The dashed line marks the random baseline (enrichment~$= 1$).}
  \label{fig:mu_enrichment_adv}
\end{figure}

\begin{enumerate}
  \item \textbf{Mixing ratio governs the advantage, not absolute density.}
    Significant advantage tracks $\mu = p_\mathrm{out}/p_\mathrm{in}$
    iso-lines, not density iso-lines, across the 27-point
    $p_\mathrm{in} \times p_\mathrm{out}$ sweep.
    Density-matched Erd\H{o}s--R\'enyi graphs yield no significant
    advantage ($p > 0.5$), confirming that community organisation,
    not density, drives the effect (Figure~\ref{fig:mu_enrichment_adv}).
    The enrichment ratio peaks at $3.87\times$ when inter-community
    edges constitute roughly 10--20\% of all edges ($\mu \approx
    0.10$--$0.15$), identifying the most favorable operating regime.

  \item \textbf{Moderate community structure.}
    Across the 27-point $p_\mathrm{in} \times p_\mathrm{out}$ sweep,
    TW2S advantage is significant ($p < 0.05$) for $\mu \in
    [0.07, 0.29]$, corresponding to modularity $Q \in [0.15, 0.41]$
    (Appendix~\ref{app:sbm_key_variable}).
    At very low $\mu$ ($< 0.07$), inter-community edges are too rare
    for Stage~1 to exploit. At high $\mu$ ($> 0.29$), community
    structure is too weak to produce identifiable bottleneck edges.

  \item \textbf{Robustness across graph size and community count.}
    The $\mu \in [0.05, 0.30]$ window is stable for $m \in \{2,3\}$
    and $n \in \{10,\ldots,32\}$. The absolute ECR advantage scales
    with $n$ as expected.
    Full results are in Appendix~\ref{app:sbm_robustness}.
\end{enumerate}

For practitioners, the applicable regime is approximately
diagnosed by a label-propagation modularity $Q \in [0.15, 0.41]$
or an inter-community edge fraction of 10--20\%.
The full parameter sweep, enrichment mechanism, and ER control
are detailed in Appendix~\ref{app:sbm_analysis}.

\subsection{From Structural Insight to Experimental Design:
            The J1--J2 TFIM Circuit}
\label{ssec:ring_motivation}
 
The structural analysis so far has used two complementary forms
of evidence: the SBM scan in Appendix~\ref{app:sbm_analysis}
characterizes TW2S behavior as a parametric function of
structural sparsity along the $\mu$ axis, while the per-family
results in Sec.~\ref{ssec:ecr_results}---in particular the
Barbell family (providing a deterministic single-bridge case
sitting at the extreme low-$\mu$ end of this axis, where TW2S
reliably identifies the unique bridge edge).
We next examine whether this characterisation transfers
to a physically motivated benchmark circuit relevant to quantum
simulation, rather than to synthetic graphs constructed for
structural analysis.
 
The J1--J2 TFIM circuit is a ring of $n$ qubits with
nearest-neighbour ($J_1$) and next-nearest-neighbour ($J_2$)
interactions.
This topology contains exactly one long-range edge, $(0, n{-}1)$,
which bridges the two ends of the ring (structurally playing
the same role as the bridge edge in the Barbell family of
Sec.~\ref{ssec:ecr_results}, but arising here from a frustrated
quantum Hamiltonian of interest in quantum simulation rather
than from a synthetic construction).
For $n = 8$, this edge has the highest normalized edge betweenness
centrality in the graph ($\mathrm{BC} = 0.1875$, compared to
$0.152$ for the next-highest edge), making it a clear structural
bottleneck.
The selection mechanism predicted by both the SBM scan and the
Barbell results (preferentially targeting the single
highest-betweenness edge that decouples the graph) therefore
applies in the same form here, and we hypothesise that TW2S
should reliably target $(0, n{-}1)$ and thereby achieve a large
ECR reduction.
 
To test this hypothesis, we compute $\Delta$ECR for TW2S and
random cut selection across $n \in \{6, 8, 10, 12\}$ and
$t \in \{1, 2, 3, 4\}$ Trotter steps (16 configurations in total),
with no noisy simulation (gate counting only).
TW2S selects $(0, n{-}1)$ in every single case
($\mathrm{top\_BC} = 1$ in all 16 configurations), confirming
that Stage~2's betweenness criterion correctly and consistently
identifies the unique bottleneck edge.
For $n = 8$ and $t \in \{1, 2, 3\}$, this yields a mean TW2S
advantage of $\Delta\mathrm{ECR} \in \{+8.7,\,+12.6,\,+15.6\}$
gates over random selection.
 
These findings provide a principled, predictive account of why
TW2S is effective on the J1--J2 TFIM circuit, and directly
motivated its choice as the primary testbed for the noise-simulation
and break-even experiments reported in
Sec.~\ref{sec:noisy_sim} and Sec.~\ref{sec:breakeven_analysis}.
Because $(0, n{-}1)$ is the unique highest-betweenness edge and
TW2S selects it with certainty, the resulting ECR reduction is
the largest attainable by any single-edge cut, a condition
that maximizes the probability of crossing the QPD break-even
threshold.
\section{Noisy Simulation on a Realistic Device Model}
\label{sec:noisy_sim}

Having characterized the conditions under which QPD is theoretically
advantageous, we now apply TW2S to the $J_1$--$J_2$ TFIM and verify
that the predicted advantage materialises in practice on the
FakeSherbrooke backend.

\subsection{Experimental Setup}
\label{sec:noisy_setup}
 
We use the one-dimensional $J_1$--$J_2$ TFIM with $J_1 = 1.0$,
$J_2 = 0.9$, $H = 1.5$, and steps~$= 4$ Trotter layers as the
benchmark circuit, targeting the strong-signal regime identified in
Section~\ref{sec:signal_strength}.
As analyzed in Sec.~\ref{ssec:ring_motivation}, the ring topology
of this circuit contains the single long-range bridging edge that
TW2S reliably targets, making it a natural testbed.
Two system sizes are tested: $n = 8$ ($|H_{\mathrm{ideal}}| = 8.38$)
and $n = 10$ ($|H_{\mathrm{ideal}}| = 9.80$).
We compare four conditions:
\begin{enumerate}
  \item \textbf{Baseline}: no gate cutting.
  \item \textbf{Random-$k$x}: randomly selected cut,
        shot multiplier $k \in \{1, 9\}$.
  \item \textbf{TW2S-$k$x}: TW2S-selected cut, same multiplier.
  \item \textbf{With M3}: additionally applying M3~\cite{m3_mitigation},
        a scalable matrix-free readout-error mitigation method.
\end{enumerate}
Each condition is repeated for five independent trials at
$M \in \{500, 2\,000, 10\,000, 50\,000\}$ shots.
All simulations run on FakeSherbrooke with the transpilation
settings of Sec.~\ref{ssec:ecr_setup}.

\subsection{ECR Reduction}
\label{sec:ecr_reduction}
 
Table~\ref{tab:ecr_counts} reports the ECR gate counts.
For both system sizes TW2S achieves a substantially larger
$\Delta\mathrm{ECR}$ than random selection, exceeding the
$\Delta\mathrm{ECR} \gtrsim 5$ threshold of
Sec.~\ref{sec:qpd_conditions} by a wide margin at $n = 8$
and remaining above it at $n = 10$ where random selection
falls below.
 
\begin{table}[h]
\centering
\caption{ECR gate counts before and after a single QPD cut.}
\label{tab:ecr_counts}
\begin{tabular}{lrrr}
\hline
                      & ECR (base) & ECR (cut) & $\Delta$ECR \\
\hline
$n=8$,  TW2S         & 297 & 250 & $\mathbf{47}$ \\
$n=8$,  Random (avg) & 297 & 288 & ${\sim}9$     \\
$n=10$, TW2S         & 335 & 321 & $\mathbf{14}$ \\
$n=10$, Random (avg) & 335 & 331 & ${\sim}4$     \\
\hline
\end{tabular}
\end{table}

\subsection{Error Reduction}
\label{sec:error_reduction}

\begin{figure*}[ht]
  \centering
  \includegraphics[width=1\linewidth]{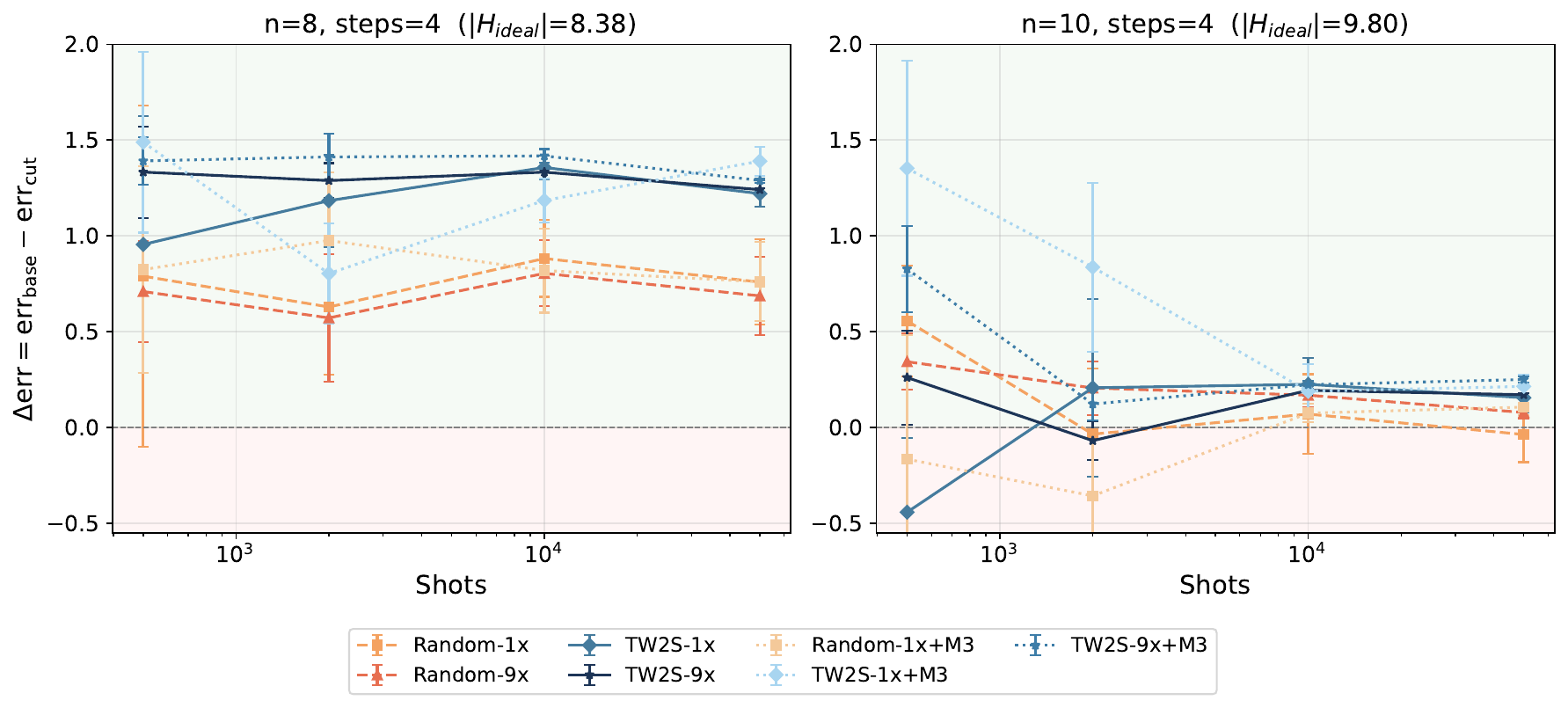}
  \caption{Error improvement $\Delta\mathrm{err} =
    \mathrm{err}_{\mathrm{base}} - \mathrm{err}_{\mathrm{cut}}$
    vs.\ shot count on FakeSherbrooke.
    Positive values indicate the cut circuit achieves lower error
    than the uncut baseline.
    Left: $n=8$; right: $n=10$.}
  \label{fig:tfim_error_vs_shots}
\end{figure*}

Figure~\ref{fig:tfim_error_vs_shots} shows
$\Delta\mathrm{err} = \mathrm{err}_{\mathrm{base}} -
\mathrm{err}_{\mathrm{cut}}$ as a function of shot count.

For $n = 8$, TW2S conditions consistently outperform the baseline
across all shot counts: TW2S-9x achieves
$\Delta\mathrm{err} \approx 1.3$ at $M = 10\,000$, corresponding to
roughly a \textbf{20\% error reduction}.
Random cutting also improves upon the baseline but by a smaller
margin ($\Delta\mathrm{err} \approx 0.8$), and the gap is maintained
across the full range of $M$.

For $n = 10$, statistical noise dominates at $M = 500$, but for
$M \ge 2\,000$ TW2S-9x and TW2S-1x consistently achieve
$\Delta\mathrm{err} \approx 0.15$--$0.2$, while Random shows
negligible or negative improvement.
This confirms the prediction of Section~\ref{sec:qpd_conditions}:
TW2S's larger $\Delta\mathrm{ECR}$ is sufficient to amortise the
mid-circuit overhead even at $n = 10$, whereas random selection's
smaller $\Delta\mathrm{ECR}$ is not.

\subsection{Effect of Readout-Error Mitigation}
\label{sec:m3_effect}

Adding M3 readout-error mitigation further amplifies the TW2S
advantage: TW2S-1x+M3 achieves $\Delta\mathrm{err} \approx 1.4$ at
$M = 50\,000$ for $n = 8$, the largest improvement observed.
In contrast, Random+M3 shows only marginal improvement over Random
without mitigation.
This suggests a practical compatibility between TW2S-based cutting and post-processing mitigation: error mitigation is most effective when the underlying circuit already has reduced noise, providing a calibration-cost-free accuracy gain for practitioners who already apply M3.
 
\section{Conclusion and future work}
\label{sec:conclusion}

We addressed the problem of routing overhead on superconducting
quantum devices, where sparse qubit connectivity forces the
insertion of additional SWAP/ECR gates when transpiling
long-range two-qubit interactions.
QPD offers a way to remove
such gates by replacing them with classical post-processing
over multiple subcircuits, but its sampling cost makes the
choice of which gate to cut critical.
Motivated by this trade-off, we proposed TW2S, a two-stage
gate-cut selection algorithm that operates entirely on
graph-structural quantities (min-fill treewidth in the first
stage and edge betweenness centrality with a degree penalty
in the second), without invoking the transpiler at selection
time.
A mean-squared-error breakeven analysis identified ECR reduction
and signal strength as the key determinants of QPD advantage.
Only ECR reduction is controllable by cut selection, which
justified TW2S as a $\Delta\mathrm{ECR}$-maximizing selector.
Empirical evaluation confirmed that TW2S outperforms random
selection when the interaction graph has identifiable sparse cuts.
Noisy simulations on FakeSherbrooke verified the predicted
advantage in practice.
Several directions remain open for future work.
Extending the algorithm to multiple simultaneous cuts would
broaden applicability while requiring careful management of the
combinatorial growth of QPD branches.
Validation on actual quantum hardware would test robustness
against time-varying calibration drift, crosstalk between
concurrently executed gates, and the queue-time cost of
dispatching multiple branch circuits, effects not captured by
static noise models.
A hybrid extension that uses backend calibration data (per-gate
fidelities, readout errors, and coherence times) to refine the
candidate cuts produced by TW2S would yield a per-backend
should-cut recommendation that is fast at the graph stage and
accurate at the final selection.
Finally, the interaction-graph abstraction collapses the temporal
structure of the circuit, and lifting the analysis to the circuit
DAG would enable gate-order-aware cut selection and joint
optimization with qubit mapping and routing in an end-to-end
compilation pipeline, where the impact of each cut on the
downstream stages is taken into account explicitly.

\bibliographystyle{ieeetr}
\bibliography{references}

@article{supreconducting,
  author    = {Morten Kjaergaard and Mollie E. Schwartz and
               Jochen Braum\"{u}ller and Philip Krantz and
               Joel I.-J. Wang and Simon Gustavsson and
               William D. Oliver},
  title     = {Superconducting Qubits: Current State of Play},
  journal   = {Annual Review of Condensed Matter Physics},
  volume    = {11},
  pages     = {369--395},
  year      = {2020},
  doi       = {10.1146/annurev-conmatphys-031119-050605},
}

@inproceedings{quantumcircuit_optimization,
  author    = {Alexander Cowtan and Silas Dilkes and Ross Duncan and
               Alexandre Krajenbrink and Will Simmons and
               Seyon Sivarajah},
  title     = {On the Qubit Routing Problem},
  booktitle = {14th Conference on the Theory of Quantum Computation,
               Communication and Cryptography (TQC 2019)},
  series    = {LIPIcs},
  volume    = {135},
  pages     = {5:1--5:32},
  year      = {2019},
  doi       = {10.4230/LIPIcs.TQC.2019.5},
}

@inproceedings{wille2019mapping,
  author    = {Robert Wille and Lukas Burgholzer and Alwin Zulehner},
  title     = {Mapping Quantum Circuits to {IBM QX} Architectures Using the Minimal Number of {SWAP} and {H} Operations},
  booktitle = {Proceedings of the 56th Annual Design Automation Conference 2019},
  series    = {DAC '19},
  pages     = {142:1--142:6},
  year      = {2019},
  publisher = {Association for Computing Machinery},
  doi       = {10.1145/3316781.3317859}
}

@article{nishio2020extracting,
  author    = {Shin Nishio and Yulu Pan and Takahiko Satoh and Hideharu Amano and Rodney Van Meter},
  title     = {Extracting Success from {IBM}'s 20-Qubit Machines Using Error-Aware Compilation},
  journal   = {ACM Journal on Emerging Technologies in Computing Systems},
  volume    = {16},
  number    = {3},
  pages     = {32:1--32:25},
  year      = {2020},
  publisher = {Association for Computing Machinery},
  doi       = {10.1145/3386162}
}

@article{Yamamoto2023VTQG,
    author = "Yamamoto, Takahiro and Ohira, Ryutaro",
    title = "{Error suppression by a virtual two-qubit gate}",
    eprint = "2212.05493",
    archivePrefix = "arXiv",
    primaryClass = "quant-ph",
    doi = "10.1063/5.0151037",
    journal = "J. Appl. Phys.",
    volume = "133",
    number = "17",
    pages = "2887757",
    year = "2023"
}

@inproceedings{Siraichi2018QubitA,
  author    = {Marcos Yukio Siraichi and Vin\'{i}cius Fernandes dos Santos
               and Caroline Collange and
               Fernando Magno Quint\~{a}o Pereira},
  title     = {Qubit Allocation},
  booktitle = {Proceedings of the 2018 International Symposium on
               Code Generation and Optimization (CGO '18)},
  pages     = {113--125},
  year      = {2018},
  publisher = {ACM},
  doi       = {10.1145/3168822},
}

@inproceedings{10.1145/3297858.3304023,
  author    = {Gushu Li and Yufei Ding and Yuan Xie},
  title     = {Tackling the Qubit Mapping Problem for {NISQ}-Era
               Quantum Devices},
  booktitle = {Proceedings of the 24th International Conference on
               Architectural Support for Programming Languages and
               Operating Systems (ASPLOS '19)},
  pages     = {1001--1014},
  year      = {2019},
  publisher = {ACM},
  doi       = {10.1145/3297858.3304023},
}

@article{Chamberland2020heavyhex,
  author  = {Christopher Chamberland and Guanyu Zhu and Theodore J. Yoder
             and Jared B. Hertzberg and Andrew W. Cross},
  title   = {Topological and Subsystem Codes on Low-Degree Graphs
             with Flag Qubits},
  journal = {Physical Review X},
  volume  = {10},
  pages   = {011022},
  year    = {2020},
  doi     = {10.1103/PhysRevX.10.011022}
}

@article{Krantz2019guide,
  author  = {Philip Krantz and Morten Kjaergaard and Fei Yan
             and Terry P. Orlando and Simon Gustavsson and William D. Oliver},
  title   = {A quantum engineer's guide to superconducting qubits},
  journal = {Applied Physics Reviews},
  volume  = {6},
  pages   = {021318},
  year    = {2019},
  doi     = {10.1063/1.5089550}
}

@article{BODLAENDER1998,
  author  = {Hans L. Bodlaender},
  title   = {A partial $k$-arboretum of graphs with bounded treewidth},
  journal = {Theoretical Computer Science},
  volume  = {209},
  number  = {1--2},
  pages   = {1--45},
  year    = {1998},
  doi     = {10.1016/S0304-3975(97)00228-4}
}

@book{NielsenChuang2010,
  author    = {Michael A. Nielsen and Isaac L. Chuang},
  title     = {Quantum Computation and Quantum Information:
               10th Anniversary Edition},
  publisher = {Cambridge University Press},
  year      = {2010},
  doi       = {10.1017/CBO9780511976667}
}

@inproceedings{Cowtan2019routing,
  author    = {Alexander Cowtan and Silas Dilkes and Ross Duncan
               and Alexandre Krajenbrink and Will Simmons
               and Seyon Sivarajah},
  title     = {On the Qubit Routing Problem},
  booktitle = {14th Conference on the Theory of Quantum Computation,
               Communication and Cryptography (TQC 2019)},
  series    = {Leibniz International Proceedings in Informatics
               (LIPIcs)},
  volume    = {135},
  pages     = {5:1--5:32},
  year      = {2019},
  doi       = {10.4230/LIPIcs.TQC.2019.5}
}

@inproceedings{Tannu2019,
  author    = {Swamit S. Tannu and Moinuddin K. Qureshi},
  title     = {Not All Qubits Are Created Equal: A Case for
               Variability-Aware Policies for {NISQ}-Era
               Quantum Computers},
  booktitle = {Proceedings of the 24th International Conference
               on Architectural Support for Programming Languages
               and Operating Systems (ASPLOS '19)},
  pages     = {987--999},
  year      = {2019},
  doi       = {10.1145/3297858.3304007}
}

@article{Javadi2024qiskit,
  author  = {Ali Javadi-Abhari and Matthew Treinish
             and Kevin Krsulich and Christopher J. Wood
             and Jake Lishman and Julien Gacon
             and Simon Martiel and Paul D. Nation
             and Lev S. Bishop and Andrew W. Cross
             and Blake R. Johnson and Jay M. Gambetta},
  title   = {Quantum Computing with {Qiskit}},
  journal = {arXiv preprint arXiv:2405.08810},
  year    = {2024},
  doi     = {10.48550/arXiv.2405.08810}
}

@article{brandes2001faster,
  author  = {Brandes, Ulrik},
  title   = {A Faster Algorithm for Betweenness Centrality},
  journal = {Journal of Mathematical Sociology},
  volume  = {25},
  number  = {2},
  pages   = {163--177},
  year    = {2001},
  doi     = {10.1080/0022250X.2001.9990249},
}

@article{peng2020,
  author    = {Tianyi Peng and Aram W. Harrow and Maris Ozols and
               Xiaodi Wu},
  title     = {Simulating Large Quantum Circuits on a Small Quantum
               Computer},
  journal   = {Physical Review Letters},
  volume    = {125},
  number    = {15},
  pages     = {150504},
  year      = {2020},
  doi       = {10.1103/PhysRevLett.125.150504},
}

@article{mitarai2021,
  author    = {Kosuke Mitarai and Keisuke Fujii},
  title     = {Constructing a Virtual Two-Qubit Gate by Sampling
               Single-Qubit Operations},
  journal   = {New Journal of Physics},
  volume    = {23},
  number    = {2},
  pages     = {023021},
  year      = {2021},
  doi       = {10.1088/1367-2630/abd7bc},
}

@article{mitarai2021overhead,
  author    = {Kosuke Mitarai and Keisuke Fujii},
  title     = {Overhead for Simulating a Non-Local Channel with Local
               Channels by Quasiprobability Sampling},
  journal   = {Quantum},
  volume    = {5},
  pages     = {388},
  year      = {2021},
  doi       = {10.22331/q-2021-01-28-388},
}

@article{ROBERTSON198449,
title = {Graph minors. III. Planar tree-width},
journal = {Journal of Combinatorial Theory, Series B},
volume = {36},
number = {1},
pages = {49-64},
year = {1984},
issn = {0095-8956},
doi = {https://doi.org/10.1016/0095-8956(84)90013-3},
url = {https://www.sciencedirect.com/science/article/pii/0095895684900133},
author = {Neil Robertson and P.D Seymour}
}

@article{ROBERTSON1986309,
title = {Graph minors. II. Algorithmic aspects of tree-width},
journal = {Journal of Algorithms},
volume = {7},
number = {3},
pages = {309-322},
year = {1986},
issn = {0196-6774},
doi = {https://doi.org/10.1016/0196-6774(86)90023-4},
url = {https://www.sciencedirect.com/science/article/pii/0196677486900234},
author = {Neil Robertson and P.D Seymour}
}

@article{doi:10.1137/0608024,
author = {Arnborg, Stefan and Corneil, Derek G. and Proskurowski, Andrzej},
title = {Complexity of Finding Embeddings in a k-Tree},
journal = {SIAM Journal on Algebraic Discrete Methods},
volume = {8},
number = {2},
pages = {277-284},
year = {1987},
doi = {10.1137/0608024},
URL = {https://doi.org/10.1137/0608024},
eprint = {https://doi.org/10.1137/0608024}
}

@article{Rose1976AlgorithmicAO,
  title={Algorithmic Aspects of Vertex Elimination on Graphs},
  author={Donald J. Rose and Robert Endre Tarjan and George S. Lueker},
  journal={SIAM J. Comput.},
  year={1976},
  volume={5},
  pages={266-283},
  url={https://api.semanticscholar.org/CorpusID:207050855}
}

@article{BODLAENDER2010259,
title = {Treewidth computations I. Upper bounds},
journal = {Information and Computation},
volume = {208},
number = {3},
pages = {259-275},
year = {2010},
issn = {0890-5401},
doi = {https://doi.org/10.1016/j.ic.2009.03.008},
url = {https://www.sciencedirect.com/science/article/pii/S0890540109000947},
author = {Hans L. Bodlaender and Arie M.C.A. Koster}
}

@article{zulehner2018efficient,
  title={An efficient methodology for mapping quantum circuits to the IBM QX architectures},
  author={Zulehner, Alwin and Paler, Alexandru and Wille, Robert},
  journal={IEEE Transactions on Computer-Aided Design of Integrated Circuits and Systems},
  volume={38},
  number={7},
  pages={1226--1236},
  year={2018},
  publisher={IEEE}
}

@article{m3_mitigation,
  title = {Scalable Mitigation of Measurement Errors on Quantum Computers},
  author = {Nation, Paul D. and Kang, Hwajung and Sundaresan, Neereja and Gambetta, Jay M.},
  journal = {PRX Quantum},
  volume = {2},
  issue = {4},
  pages = {040326},
  numpages = {9},
  year = {2021},
  month = {Nov},
  publisher = {American Physical Society},
  doi = {10.1103/PRXQuantum.2.040326},
  url = {https://link.aps.org/doi/10.1103/PRXQuantum.2.040326}
}

@article{pfeuty1970,
  author  = {Pfeuty, Pierre},
  title   = {The one-dimensional Ising model with a transverse field},
  journal = {Annals of Physics},
  volume  = {57},
  number  = {1},
  pages   = {79--90},
  year    = {1970},
  doi     = {10.1016/0003-4916(70)90270-8}
}

@book{sachdev2011,
  author    = {Sachdev, Subir},
  title     = {Quantum Phase Transitions},
  edition   = {2},
  publisher = {Cambridge University Press},
  year      = {2011},
  doi       = {10.1017/CBO9780511973765}
}
\appendices
 
\section{Experimental Parameters for the Failure-Mode Analysis}
\label{app:failure_mode_params}

The failure-mode experiments of Sec.~\ref{sec:failure_modes} use
the one-dimensional transverse-field Ising model (1D TFIM)
\cite{pfeuty1970,sachdev2011} as the benchmark circuit.
The Hamiltonian is
\begin{equation}
  H \;=\; J_1 \sum_{\langle i,j\rangle} Z_i Z_j
        + h \sum_{i} X_i,
  \label{eq:tfim_hamiltonian}
\end{equation}
realised as a fixed-angle Trotterisation with parameters
$J_1 = 1.0$, $h = 0.7$, rzz angle $\varphi_{J_1} = \pi/2$,
single-qubit $x$-rotation step $dt_x = 0.1$, and initial state
$|+\rangle^{\otimes n}$.
The observable estimated is the Hamiltonian $H$ itself. The ideal
value $\langle H\rangle_\mathrm{ideal}$ is obtained by exact
statevector simulation.

\paragraph{Configuration grid.}
We sweep 18 configurations spanning $n \in \{4, 6, 8, 10\}$ with
Trotter steps $T \in \{1, 2, 3, 4\}$, and $n = 12$ restricted to
$T \in \{2, 4\}$ to bound classical-simulation cost.
Seven shot budgets are evaluated:
$M \in \{100, 300, 1\mathrm{K}, 3\mathrm{K}, 10\mathrm{K},
30\mathrm{K}, 100\mathrm{K}\}$.
Two shot-allocation strategies are compared: shared budget
($\times 1$, where all 6 branch circuits share the total $M$
shots) and per-subcircuit ($\times 1.5$, i.e.\ $\times 9$ total).
Each configuration is repeated $R = 5$ times with independent
simulator seeds.

\paragraph{Cut selection and transpilation.}
Because the failure-mode analysis aims to characterize when QPD
is beneficial rather than to evaluate TW2S itself, we use an
oracle-style selection: Stage~1 of TW2S
(Sec.~\ref{subsec:bottleneck-ranking}) proposes a shortlist of
$K=3$ candidate edges, and the final cut is the candidate
achieving the largest post-transpilation $\Delta\mathrm{ECR}$
on \textsc{FakeSherbrooke}.
This maximizes the ECR reduction for each configuration,
isolating the QPD payoff structure from the cut-selection method.
All transpilations use Qiskit \texttt{optimization\_level}$\,=1$
with a fixed transpiler seed.

\section{Detailed Derivation of the QPD Breakeven Condition}
\label{app:theoretical_breakeven}
 
This appendix provides the full derivation of the idealized
breakeven analysis summarised in Sec.~\ref{sec:breakeven_analysis}.
Under the idealized model of Sec.~\ref{sec:breakeven_analysis}
(assumptions (i)--(iv)), the deviations from this criterion
observed on realistic hardware are characterized in
Sec.~\ref{sec:failure_modes}.
Throughout this appendix $\Delta N$ denotes the post-cut ECR-count
reduction (i.e.\ $\Delta\mathrm{ECR}$, as defined in
Sec.~\ref{sec:breakeven_analysis}).
 
\paragraph{The $\times 9$ case: unconditional advantage in the idealized model.}
The QPD variance penalty per cut is $\gamma^2 = 9$
(Sec.~\ref{subsec:qpd}).
When each branch circuit receives $1.5\times$ the baseline shots
(total budget $9\times$), the variance penalty is exactly cancelled:
\begin{equation}
  \frac{\gamma^2}{m} = \frac{9}{9} = 1.
\end{equation}
Under the idealized depolarizing-bias model adopted in this
appendix, gate cutting simultaneously reduces the
systematic noise bias for any $\Delta N > 0$, so the
$\times 9$ allocation is advantageous whenever
$\Delta N > 0$ and no breakeven curve is needed.
Sec.~\ref{sec:failure_modes} discusses the conditions under
which this idealized guarantee fails on realistic noisy hardware.
 
\paragraph{The $\times 1$ case: equal total shot budget.}
With $m = 1$, the variance penalty $\gamma^2 = 9$ is fully incurred.
Modelling MSE as the sum of squared bias and variance,
\begin{align}
  \mathrm{MSE}_{\mathrm{base}}
    &= \mathrm{bias}^2(N) + \frac{\sigma_H^2}{M}, \\
  \mathrm{MSE}_{\mathrm{QPD}}
    &= \mathrm{bias}^2(N-\Delta N) + \frac{\gamma^2 \sigma_H^2}{M},
\end{align}
with depolarising bias
$\mathrm{bias}(N) \approx
|\langle H\rangle_{\mathrm{ideal}}|(1 - e^{-pN})$,
where $p$ is the per-gate ECR error rate and $N$ is the ECR gate
count of the baseline circuit.
Gate cutting reduces the ECR count from $N$ to $N-\Delta N$,
decreasing the bias at the cost of inflating the variance by
$\gamma^2$.
 
Setting $\mathrm{MSE}_{\mathrm{QPD}} < \mathrm{MSE}_{\mathrm{base}}$
and substituting the bias approximation gives
\begin{align}
  &|\langle H\rangle_{\mathrm{ideal}}|^2
   \left(1-e^{-p(N-\Delta N)}\right)^2
   + \frac{\gamma^2 \sigma_H^2}{M} \notag \\
  &\quad <\;
   |\langle H\rangle_{\mathrm{ideal}}|^2
   \left(1-e^{-pN}\right)^2
   + \frac{\sigma_H^2}{M}.
\end{align}
Rearranging to isolate $M$,
\begin{align}
  &\frac{(\gamma^2 - 1)\,\sigma_H^2}{M} \notag \\
  &\quad <\;
  |\langle H\rangle_{\mathrm{ideal}}|^2
  \!\left[
    \left(1-e^{-pN}\right)^2
    - \left(1-e^{-p(N-\Delta N)}\right)^2
  \right],
\end{align}
which yields the breakeven shot count
\begin{equation}
  M^*(\Delta N)
  = \frac{(\gamma^2-1)\,\sigma_H^2}
         {|\langle H\rangle_{\mathrm{ideal}}|^2
          \!\left[\left(1-e^{-pN}\right)^2
                 -\left(1-e^{-p(N-\Delta N)}\right)^2\right]},
  \label{eq:Mstar_app}
\end{equation}
identical to Eq.~\eqref{eq:Mstar} of the main text.
QPD with equal total budget wins if and only if $M > M^*(\Delta N)$.
Note that the denominator is positive whenever $\Delta N > 0$,
since $e^{-pN} < e^{-p(N-\Delta N)}$ implies
$(1-e^{-pN})^2 > (1-e^{-p(N-\Delta N)})^2$.
 
The three key levers in Eq.~\eqref{eq:Mstar_app} are:
\begin{enumerate}
  \item \textbf{$\Delta N$}: the primary output of
        TW2S. Larger reduction drives $M^*$ down.
  \item \textbf{$|\langle H\rangle_{\mathrm{ideal}}|$}: lowers $M^*$
        quadratically. As it approaches zero, $M^* \to \infty$ and
        QPD cannot win at any finite shot count.
  \item \textbf{$pN$}: most favourable near unity (circuits deep
        enough to accumulate significant but not saturated noise).
\end{enumerate}
 
\begin{figure}[ht]
  \centering
  \includegraphics[width=1\linewidth]{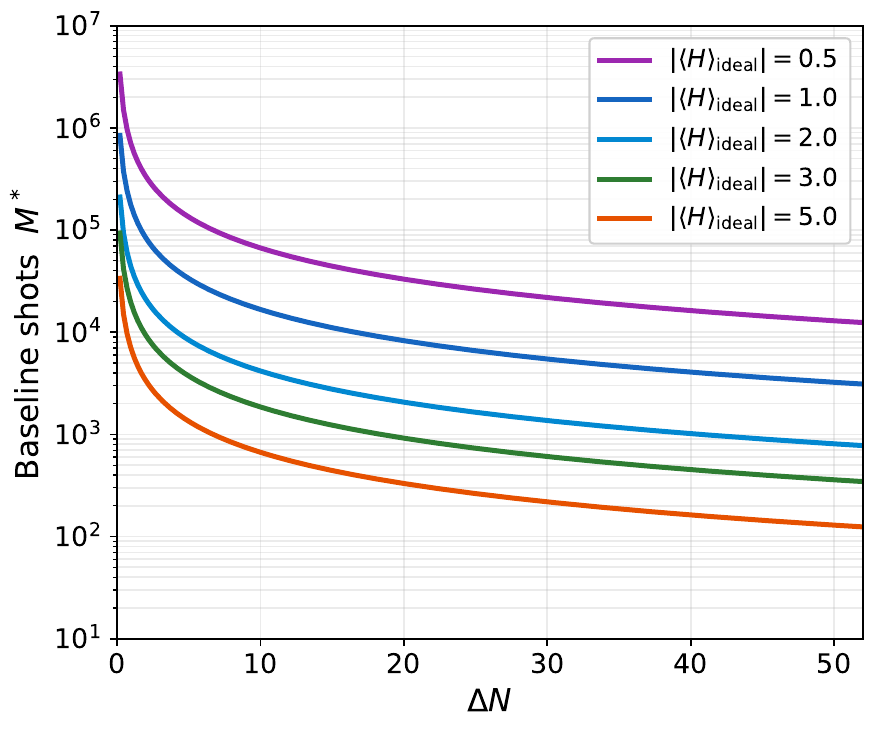}
  \caption{Theoretical breakeven shot count $M^*$
    (Eq.~\eqref{eq:Mstar_app}, $p=0.005$, $\sigma_H=7$, $N=200$).
    QPD with equal budget wins \emph{above} each curve.
    Larger $|\langle H\rangle_{\mathrm{ideal}}|$ shifts curves
    downward, reducing the required shot count.}
  \label{fig:breakeven_H0}
\end{figure}

Figure~\ref{fig:breakeven_H0} visualises $M^*(\Delta N)$ for
$p = 0.005$, $N = 200$.
For $|\langle H\rangle_{\mathrm{ideal}}| \ge 5$ (achieved at
steps$\,{=}\,4$, $n \ge 8$), the threshold drops below $10^3$ shots
even at moderate $\Delta N\approx 10$--$15$.
 
\section{Computational Complexity of TW2S}
\label{app:complexity}
 
This appendix details the per-step computational cost of the
two-stage cut selection algorithm
(Algorithm~\ref{alg:two-stage}).
Let $n=|V_I|$ denote the number of logical qubits and
$m=|E_I|$ the number of edges in the interaction graph.
Table~\ref{tab:complexity} summarizes the cost of each
step.
 
\begin{table}[htbp]
\centering
\caption{Per-step computational cost of the two-stage
         cut selection algorithm.}
\label{tab:complexity}
\begin{tabular}{lll}
\hline
Step & Operation & Cost \\
\hline
1 & Interaction graph construction & $O(|G_{\mathrm{2Q}}|)$ \\
2 & Min-fill trace (Stage~1) & $O(n^3)$ worst case \\
3 & Edge score accumulation & $O(nm)$ \\
4 & Edge betweenness centrality (Stage~2) & $O(nm)$ \\
5 & Score maximization over $K$ candidates & $O(K)$ \\
\hline
\textbf{Total} & & $O(n^3 + nm)$ \\
\hline
\end{tabular}
\end{table}
 
\paragraph{Min-fill trace.}
The min-fill heuristic eliminates vertices one by one,
each time choosing the vertex that introduces the fewest
fill edges.
Computing the fill count for a single vertex $v$ costs
$O(\deg(v)^2)$. Summing over all $n$ elimination steps
gives $O(n\Delta^2)$ where $\Delta$ is the maximum degree,
which is $O(n^3)$ in the worst case for dense graphs.
For the sparse interaction graphs arising from practical
quantum circuits, the actual runtime is substantially lower.
 
\paragraph{Edge betweenness centrality.}
We compute betweenness centrality using Brandes'
algorithm~\cite{brandes2001faster}, which runs a
breadth-first search from each vertex and accumulates
dependency scores in $\mathcal{O}(n+m)$ per source.
The total cost is $\mathcal{O}(n(n+m))$, which reduces to $\mathcal{O}(nm)$
for sparse graphs ($m \ll n^2$) and to $\mathcal{O}(n^2)$ when
$m = \mathcal{O}(n)$ (e.g., chain or tree topologies).
 
\paragraph{Comparison with exhaustive evaluation.}
A naive approach that evaluates the post-cut ECR count
for every candidate edge would require $|E_I|$
transpilation calls, each of which may take several
seconds on a modern workstation.
Our method replaces all transpilation calls with $O(n^3+nm)$
graph computations, reducing selection overhead to
sub-millisecond latency for circuits with up to a few
hundred qubits.

\section{Per-Family ECR Reduction Details}
\label{app:ecr_per_graph}

This appendix explains the mechanism behind TW2S performance for
the three non-SBM graph families (Grid, Watts--Strogatz, Barbell).
Numerical results are in Table~\ref{tab:perf_summary}.

\paragraph{Grid graphs: positive but not significant overall.}
TW2S achieves a positive mean advantage ($+2.2$ ECR) across all
16 configurations, but the result does not reach statistical
significance ($p = 0.295$) over the full $3{\times}3$--$6{\times}6$
range.
The advantage is driven by larger grids: for $5{\times}5$ and
$6{\times}5$ configurations the individual advantages are
substantially larger, but variance across seeds is high for
small grids, pulling the aggregate $p$-value above threshold.
On grid graphs, Stage~1 already identifies the geometric
separator edge as the treewidth bottleneck, and Stage~2's
betweenness-centrality re-ranking provides no additional benefit
on average (oracle efficiency 0.35 vs 0.37 for Stage~1 alone,
as detailed in Appendix~\ref{app:two_stages}).
The TW2S advantage on grids therefore derives primarily from
Stage~1's treewidth-guided shortlisting.

\paragraph{Watts--Strogatz graphs: advantage concentrated at intermediate degree.}
At $k = 4$, the graph has enough intra-cluster structure to
produce identifiable bridge edges that Stage~1 can surface,
yielding 85\% win rate (mean $+6.4$ ECR, $p < 0.001$).
At $k = 6$, bridge structure weakens and TW2S shows no
significant advantage ($p = 0.382$).
At $k = 2$, the graph approaches a sparse cycle. TW2S achieves
a statistically significant positive advantage ($p = 0.029$,
mean $+6.9$ ECR) but only 50\% win rate, reflecting high
variance across seeds.

\paragraph{Barbell graphs: advantage reverses at large clique sizes.}
TW2S reliably identifies the bridge edge for small cliques
($k \in \{3, 4\}$, 100\% win rate, $p < 0.05$).
For large cliques ($k \geq 5$), however, TW2S yields negative
advantage: at $k = 5$ mean $-3.3$ ECR (0\% win, $p = 0.011$),
and at $k = 8$ mean $-10.2$ ECR (25\% win).
This reversal occurs because for large cliques the bridge edge
connecting the two cliques is no longer the dominant routing
bottleneck: each clique's dense internal structure generates
substantial intra-clique SWAP overhead, and cutting the bridge
edge leaves this overhead unchanged while adding QPD sampling
cost.
Stage~1's treewidth score correctly surfaces the bridge as a
sparse cut, but the bridge's betweenness centrality also
becomes relatively lower as intra-clique paths accumulate,
making Stage~2's selection less reliable in this regime.

\section{Marginal Contribution of Stage~2}
\label{app:two_stages}

This appendix quantifies the marginal contribution of Stage~2
beyond Stage~1 alone, using the same 240-instance evaluation set
(Grid, Watts--Strogatz, Barbell, and SBM families), per-edge CNOT
circuit construction, and shortlist size $K=3$ as
Sec.~\ref{sec:algo_verification}.
For each instance, exhaustive oracle evaluation computes
$\Delta\mathrm{ECR}$ for every possible cut edge. We define
\emph{oracle efficiency} as the fraction of the maximum possible
$\Delta\mathrm{ECR}$ captured by each strategy.

\paragraph{Results.}
Stage~1 top-1 alone achieves a mean oracle efficiency of 0.37,
the full TW2S pipeline (Stage~1+2) achieves 0.45, and random
selection achieves 0.28.
Stage~2 changes Stage~1's selection in 134 of 240 instances
(55.8\%). Of these, Stage~2 improves upon Stage~1 in 75 cases
(56\%), is neutral in 18, and is counterproductive in 41.
The mean s2 benefit across all changed cases is $+1.3$ ECR.

\paragraph{Family-level breakdown.}
The contribution of Stage~2 varies substantially by graph family
(Table~\ref{tab:stage_decomp}).
Stage~2 provides the largest gain on SBM graphs (1.24$\times$) and
Barbell graphs (2.04$\times$), where inter-community bridge edges
and clique-bridge edges are globally central and betweenness
centrality correctly identifies them.
On WS graphs Stage~2 adds modest improvement (1.06$\times$).
On Grid graphs Stage~2 slightly degrades Stage~1 performance
(0.92$\times$): in regular grid topologies, the treewidth
bottleneck identified by Stage~1 already corresponds to the
geometric separator, and betweenness centrality re-ranks
candidates in a way that does not improve ECR reduction on average.

\begin{table}[h]
\centering
\caption{Oracle efficiency by family and stage ($K=3$,
         per-edge CNOT, 240 instances total).}
\label{tab:stage_decomp}
\begin{tabular}{lrrrr}
\hline
Family & Stage-1 & TW2S & Random & TW2S/S1 \\
\hline
Grid    & 0.37 & 0.35 & 0.26 & 0.92$\times$ \\
WS      & 0.38 & 0.40 & 0.27 & 1.06$\times$ \\
Barbell & 0.15 & 0.30 & 0.36 & 2.04$\times$ \\
SBM     & 0.41 & 0.51 & 0.27 & 1.24$\times$ \\
\hline
Overall & 0.37 & 0.45 & 0.28 & 1.20$\times$ \\
\hline
\end{tabular}
\end{table}

\paragraph{Interpretation.}
The overall $1.20\times$ improvement confirms that Stage~2 adds
value on average, but the family-level results reveal that its
contribution is concentrated in graphs with identifiable
inter-community or bridge structure (SBM and Barbell), where
betweenness centrality correctly surfaces the globally central
bottleneck edge from Stage~1's shortlist.
In graphs with regular geometry (Grid), Stage~1 already
identifies the routing bottleneck and Stage~2's re-ranking
provides no additional benefit.
The mechanism of Stage~2 is described in
Sec.~\ref{subsec:structural-tiebreaker}.

\section{Structural Analysis of the SBM Mixing-Ratio Dependence}
\label{app:sbm_analysis}

This appendix provides the full structural analysis of the SBM
mixing-ratio dependence summarised in
Sec.~\ref{ssec:tw2s_conditions}.
The primary metric throughout is the \emph{Stage-1 top-1
advantage} (TW-1): the ECR reduction achieved by the single edge
selected by Stage~1 alone, relative to a random-cut baseline.
We use TW-1 rather than full TW2S throughout this appendix
in order to isolate Stage~1's behavior from Stage~2's
re-ranking. The combined TW2S results are reported in
Sec.~\ref{sec:algo_verification}.

\subsection{Identifying the Mixing Ratio as the Key Variable}
\label{app:sbm_key_variable}

We adopt SBM rather than deterministic single-bridge topologies
such as Barbell for this structural analysis because SBM admits
a continuous parameter $\mu = p_\mathrm{out}/p_\mathrm{in}$ that
smoothly interpolates between a trivially partitioned graph
($\mu \to 0$) and an Erd\H{o}s--R\'enyi-like graph ($\mu \to 1$).
This allows us to resolve the \emph{transition} between
``TW2S advantage'' and ``no advantage'' as a function of
structural strength, a transition that cannot be observed on
deterministic sparse-cut topologies.
Furthermore, SBM admits a density-matched ER comparison, which
we exploit to disentangle the effect of community structure from
that of absolute graph density.

To determine which structural parameters govern Stage-1
effectiveness, we conduct a broad two-dimensional sweep over the
SBM parameters $p_\mathrm{in}$ (intra-community edge probability)
and $p_\mathrm{out}$ (inter-community edge probability).
We fix $n = 16$, $m = 2$ and vary $p_\mathrm{in} \in
\{0.3, 0.4, 0.5, 0.6, 0.7\}$ and $p_\mathrm{out} \in
\{0.02, 0.05, 0.10, 0.20, 0.30, 0.40\}$ with $p_\mathrm{out} <
p_\mathrm{in}$, yielding 27 grid points evaluated over 20 seeds
each.

\begin{figure}[ht]
  \centering
  \includegraphics[width=1\linewidth]{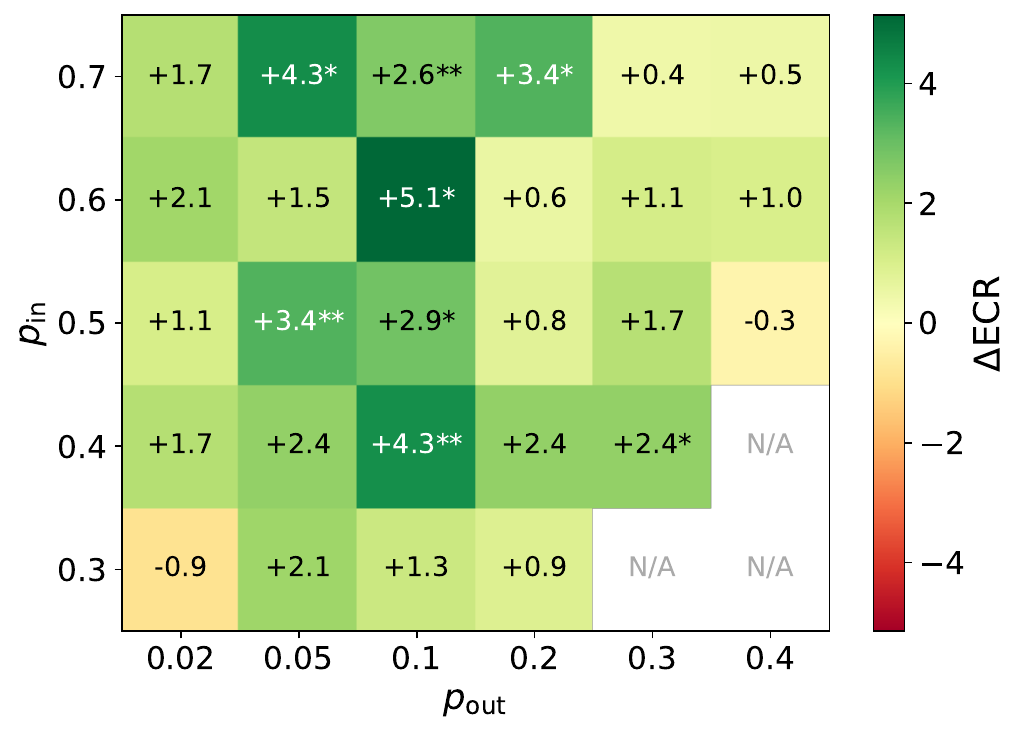}
  \caption{Mean TW-1 advantage ($\Delta$ECR gates) across the
           $p_\mathrm{in} \times p_\mathrm{out}$ grid
           ($n=16$, $m=2$, 20 seeds per cell).
           Cells annotated with $**$ ($p < 0.01$) or $*$ ($p < 0.05$)
           are statistically significant.
           The significant region follows diagonal lines of constant
           $\mu = p_\mathrm{out}/p_\mathrm{in}$, not horizontal or
           vertical density iso-lines, identifying $\mu$ as the key
           structural variable.}
  \label{fig:pin_pout_heatmap}
\end{figure}

Figure~\ref{fig:pin_pout_heatmap} shows the mean TW-1 advantage
for each cell.
The most striking feature is the \emph{diagonal structure} of the
significant cells: 7 of the 8 statistically significant cells
($p < 0.05$) lie in the range $\mu \in [0.07, 0.29]$, spanning
graph densities from 0.25 to 0.45.
These significant cells do not cluster along rows (constant
$p_\mathrm{in}$) or columns (constant $p_\mathrm{out}$) of the
grid, but instead follow lines of constant
$\mu = p_\mathrm{out}/p_\mathrm{in}$.
By contrast, only 1 of the 12 cells with $\mu > 0.30$ reaches
significance, regardless of the absolute density.
This pattern directly implies that the \emph{mixing ratio}
$\mu$ (not $p_\mathrm{in}$ or $p_\mathrm{out}$ individually,
and not the overall graph density) is the operative variable.

\paragraph{Enrichment mechanism.}
Having identified $\mu$ as the key variable from the
$p_\mathrm{in} \times p_\mathrm{out}$ sweep, we now analyze
\emph{why} $\mu$ governs Stage-1 effectiveness using a dedicated
$\mu$-sweep experiment ($p_\mathrm{in} = 0.5$ fixed, $\mu \in
\{0.02, 0.05, \ldots, 0.40\}$, 20 seeds each).
Let $r_\mathrm{inter}$ denote the fraction of all edges that cross
community boundaries, and $\hat{r}_\mathrm{inter}$ the fraction
of trials in which Stage~1 selects an inter-community edge.
We define the \emph{enrichment ratio} as
\begin{equation}
  \mathrm{Enrichment} = \frac{\hat{r}_\mathrm{inter}}{r_\mathrm{inter}},
  \label{eq:enrichment}
\end{equation}
which equals 1 when Stage~1 selects inter-community edges at the
same rate as random, and exceeds 1 when it preferentially targets
them.
For the ER controlled comparison below, we additionally evaluate
SBM graphs at $\mu \in \{0.50, 0.70\}$ ($n = 16$, $m = 2$,
20 seeds each); these high-$\mu$ points lie outside the main
$\mu$-sweep window and are used solely to enable density-matched
ER comparison.

At $\mu = 0.10$, Stage~1 selects inter-community edges at
$3.87\times$ the rate of a random selector, and achieves a
statistically significant advantage (mean $+3.36$ ECR gates,
$p = 0.001$, win rate 80\%).
As $\mu$ increases, the enrichment ratio falls monotonically toward
1, and the advantage loses statistical significance beyond
$\mu \approx 0.20$ in this fixed-$p_\mathrm{in}$ sweep.
At $\mu = 0.02$, despite strong community structure ($Q = 0.46$),
almost no inter-community edges exist ($r_\mathrm{inter} = 0.02$),
so Stage~1 is forced to select intra-community edges and the
advantage is negligible.

The mechanism is thus clear. Stage~1 preferentially surfaces
inter-community bridge edges, and this enrichment is strongest
when $\mu$ is small enough that such edges are structurally
distinctive.
To confirm that enrichment directly mediates the advantage, we
condition on the type of edge Stage~1 selects.
Across $\mu \in [0.05, 0.30]$ and all seeds, the mean advantage when Stage~1
selects an inter-community edge is $+4.48$ ECR gates ($n = 47$),
compared to $+1.32$ ECR gates when it selects an intra-community
edge ($n = 53$; two-sample $t$-test $p = 0.002$).
Because Stage~2 operates exclusively on Stage~1's candidate
shortlist, the Stage-1 enrichment characterized here defines the
\emph{ceiling} of selection quality available to the full TW2S
pipeline.

\subsection{Controlled Comparison: Community Structure vs.\ Density}
\label{app:sbm_er_control}

The $\mu$-sweep and $p_\mathrm{in} \times p_\mathrm{out}$ grid
are both based on SBM graphs, where $\mu$ and density are
correlated by construction.
To disentangle these effects, we compare SBM graphs against
Erd\H{o}s--R\'enyi (ER) random graphs matched for density.

An ER graph $G(n, p_\mathrm{ER})$ has no community organisation by
construction.
We generate ER graphs with $n = 16$ and $p_\mathrm{ER} \in
\{0.376, 0.444\}$, matching the mean densities of SBM $\mu = 0.50$
and $\mu = 0.70$ respectively, over 20 seeds each.
At matched density, ER graphs yield no statistically significant
advantage ($p > 0.5$ at both density levels), while the
corresponding SBM graphs retain advantage
($\mu = 0.50$: mean $+2.5$ ECR, $p = 0.012$;
 $\mu = 0.70$: mean $+4.8$ ECR, $p < 0.001$).
This pattern supports the interpretation that the observed
advantage reflects community organisation rather than graph
density alone.

A residual advantage persists in the high-$\mu$ SBM regime
($\mu \geq 0.50$, where enrichment $\approx 1$ and community
structure has largely collapsed to $Q \approx 0.01$).
We hypothesize that this residual reflects local density
heterogeneity intrinsic to the SBM generative process, since it
does not appear in the density-matched ER controls. A definitive
mechanistic explanation is left for future work.

\subsection{Robustness across Graph Size \texorpdfstring{$n$}{n}
            and Community Count \texorpdfstring{$m$}{m}}
\label{app:sbm_robustness}

\label{app:robustness_nm}

These experiments probe the robustness of the Stage-1
$\mu$-dependence using TW-1 (i.e., the single edge selected by
Stage~1 alone), not the full TW2S pipeline.
Two experiments verify that the $\mu$-governed Stage-1 advantage
identified in Sec.~\ref{ssec:tw2s_conditions} generalises beyond
the $n=16$, $m=2$ reference configuration:
(i) varying $m$ at fixed $n=24$, and (ii) varying $n$ at fixed
$m=2$.
The takeaway is threefold: (a) for $m \in \{2, 3\}$ the
$\mu$-window is approximately preserved ($[0.10, 0.20]$ for $m=2$,
$[0.05, 0.30]$ for $m=3$). (b) At $m=4$ the
single-cut budget cannot resolve four community boundaries and
the $\mu$-dependence breaks down. (c) Absolute Stage-1 advantage
grows with $n$, as expected from the larger gate count.

\begin{figure*}[ht]
  \centering
  \includegraphics[width=\linewidth]{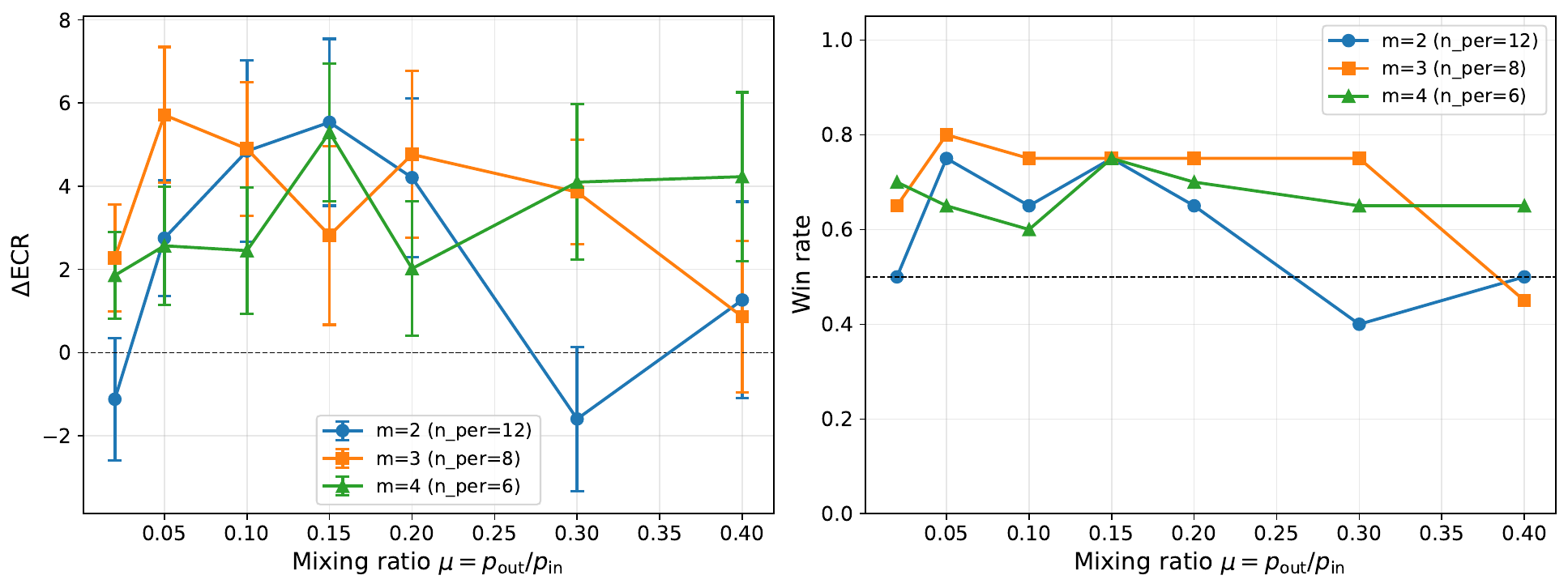}
  \caption{TW-1 advantage and win rate vs.\ $\mu$ for $n=24$
           fixed, $m \in \{2,3,4\}$ (20 seeds per cell).
           Significance markers: $**$ $p<0.01$, $*$ $p<0.05$.}
  \label{fig:fix_n_vary_m}
\end{figure*}

\paragraph{Varying $m$ at fixed $n=24$.}
We set $n=24$ and vary $m \in \{2,3,4\}$ ($n_\mathrm{per} \in
\{12,8,6\}$), sweeping $\mu$ over 7 values with 20 seeds each
(Figure~\ref{fig:fix_n_vary_m}).
For $m=2$, significant advantage ($p<0.05$) is
concentrated in $\mu \in [0.10, 0.20]$; for $m=3$, the significant
range extends to $\mu \in [0.05, 0.30]$.
Both are broadly consistent with the main result, confirming that
the $\mu$-window is approximately preserved across community counts.
For $m=4$, only scattered cells reach significance. With four
community boundaries and a single-cut budget, Stage~1 cannot
reliably identify a dominant bridge edge.

\begin{figure*}[ht]
  \centering
  \includegraphics[width=\linewidth]{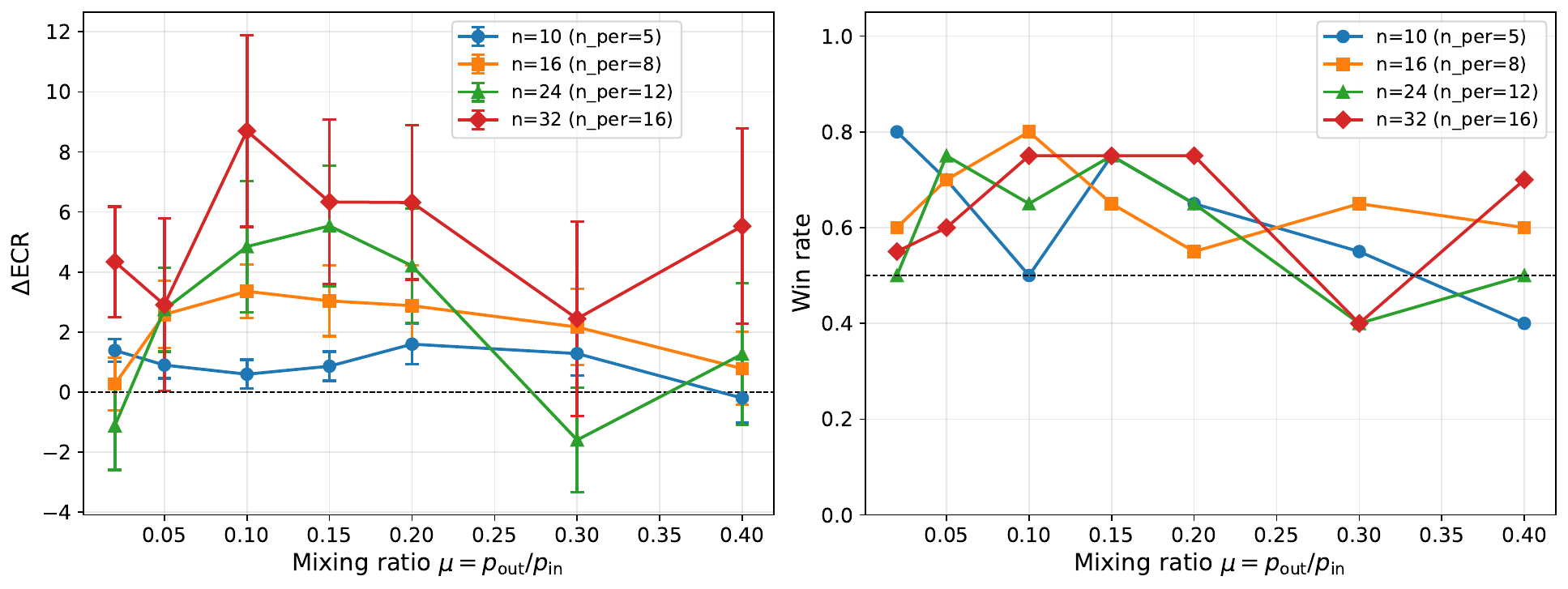}
  \caption{TW-1 advantage and win rate vs.\ $\mu$ for $m=2$
           fixed, $n \in \{10,16,24,32\}$ (20 seeds per cell).}
  \label{fig:fix_m_vary_n}
\end{figure*}

\paragraph{Varying $n$ at fixed $m=2$.}
We set $m=2$ and vary $n \in \{10,16,24,32\}$ ($n_\mathrm{per}
\in \{5,8,12,16\}$), with data for $n \in \{10,16\}$ reused
from the main robustness dataset
(Figure~\ref{fig:fix_m_vary_n}).
The significant $\mu$ range ($[0.05, 0.20]$) is stable across
all four sizes. Only the absolute advantage grows with $n$
(peak $+1.6$ ECR at $n=10$ to $+8.7$ ECR at $n=32$), as
expected from the larger gate counts.

\EOD

\end{document}